\documentclass[authoryear,3p,times]{elsarticle}

\usepackage[utf8]{inputenc}
\usepackage[T1]{fontenc}
\usepackage{url}
\usepackage{booktabs}
\usepackage{amsmath}
\usepackage{amsthm}
\usepackage{amsfonts}
\usepackage{nicefrac}
\usepackage{microtype}
\usepackage{xcolor}
\usepackage{pifont}
\usepackage{multirow}
\usepackage{graphicx}
\usepackage[textfont=it,labelfont=bf]{caption}
\usepackage{enumitem}
\usepackage{array}
\usepackage{graphbox}
\usepackage{subcaption}
\usepackage{placeins}
\usepackage{tikz}
\usepackage{pgfplots}
\pgfplotsset{compat=1.17}
\usetikzlibrary{arrows.meta,calc,positioning}
\usepackage{hyperref}
\hypersetup{
    breaklinks=true,
    colorlinks,
    linkcolor={blue!80!black},
    citecolor={blue!80!black},
    urlcolor={blue!80!black},
    plainpages=true
}
\usepackage{bm}

\newcommand{\xref}[2]{\hyperref[#2]{#1~\ref*{#2}}}
\newcommand{\colref}[3]{\hyperref[#2]{#1~\ref*{#2}{#3}}}
\newcommand{\figref}[1]{\xref{Figure}{#1}}

\newcommand{\secref}[1]{\xref{Section}{#1}}
\newcommand{\eqnref}[1]{\xref{Equation}{#1}}

\newcommand{\pvzoompanel}[1]{%
    \includegraphics[width=0.88\linewidth,trim=250 655 230 18,clip]{#1}\\[-0.25em]%
    \includegraphics[width=0.96\linewidth,trim=235 105 235 105,clip]{#1}%
}
\definecolor{mplblue}{RGB}{31,119,180}
\definecolor{mplorange}{RGB}{255,127,14}
\newcommand{\distpanel}[1]{%
    \includegraphics[width=\linewidth]{#1}%
}
\newcommand{\distlegend}{%
    {\small
    \raisebox{0.25ex}{\textcolor{mplblue}{\rule{1.35em}{0.75ex}}}\hspace{0.35em}$E_{\mathrm{eff}}$
    \hspace{1.8em}
    \raisebox{0.25ex}{\textcolor{mplorange}{\rule{1.35em}{0.75ex}}}\hspace{0.35em}$\nu_{\mathrm{eff}}$}%
}
\newcommand{\macropanel}[1]{%
    \includegraphics[width=\linewidth]{#1_cbar.pdf}\\[0.15em]%
    \includegraphics[width=\linewidth]{#1_figure.pdf}%
}

\begin{document}

\begin{frontmatter}

\title{From Sparse AFM Observations to Probabilistic Macroscale Mechanics: A Generative-Physics Framework for Plant Cell Walls}

\author[me]{Samundra Karki}
\ead{samundra@iastate.edu}

\author[ece]{Angona Biswas}
\ead{angona3@iastate.edu}

\author[ece]{Nabila Masud}
\ead{nabila@iastate.edu}

\author[ece]{Md Hasibul Hasan Hasib}
\ead{hasib@iastate.edu}

\author[ece]{Anwesha Sarkar\corref{cor}}
\ead{anweshas@iastate.edu}

\author[me,trac,psi]{Baskar Ganapathysubramanian\corref{cor}}
\ead{baskarg@iastate.edu}

\cortext[cor]{Corresponding authors}

\address[me]{Department of Mechanical Engineering, Iowa State University, Ames, Iowa 50011, USA}
\address[ece]{Department of Electrical and Computer Engineering, Iowa State University, 2520 Osborn Drive, Ames, IA 50011, USA}
\address[trac]{Translational AI Research and Education Center, Iowa State University, Ames, Iowa 50011, USA}
\address[psi]{Plant Science Institute, Iowa State University, Ames, Iowa 50011, USA}

\begin{abstract}
Nanoscale imaging of developing plant cell walls is expensive, and a limited number of atomic force microscopy (AFM) scans cannot capture the full structural variability of the wall.
We present a generative-physics workflow that links sparse AFM observations of cotton (\textit{Gossypium hirsutum}) fiber cell walls at 8 days post-anthesis (DPA) to distributions of effective elastic properties and a larger-scale mechanical response.
We adapt a Stable Diffusion model with Low-Rank Adaptation (LoRA) to expand the experimental scans into an ensemble of AFM-like microstructures and assess the synthetic structures using microfibril crossover count and crossover angle.
Each microstructure is mapped to spatially varying Young's modulus and Poisson's ratio fields and analyzed using strain-controlled finite element homogenization.
Repeating this process across the image ensemble and a prescribed sweep of matrix-to-fibril stiffness ratios produces distributions of effective Young's modulus and Poisson's ratio. A single constitutive pair cannot capture this variation.
The resulting distributions depend strongly on the matrix-to-fibril stiffness ratio and, in several cases, exhibit apparent multimodality.  
We further propagate the paired effective properties into 20 stochastic realizations of a tensile simulation of a larger specimen, yielding a distribution of macroscale stress response.
The framework provides a probabilistic link between sparse nanoscale observations and continuum-scale mechanics while retaining variability across scales.
The present results establish the computational workflow, while calibration of the intensity-to-property mapping against nanomechanical measurements remains necessary for quantitative prediction at the fiber scale.
\end{abstract}


\end{frontmatter}

\section{Introduction}
\label{sec:introduction}

The mechanical behavior of a plant cell wall emerges from its organization across multiple length scales.
In developing cotton fibers, the nanoscale arrangement of cellulose microfibrils within the pectin- and hemicellulose-rich matrix governs the effective response observed at the fiber scale \citep{kim2001cotton,haigler2012cotton}.
Atomic force microscopy (AFM) can resolve this nanoscale morphology, together with the local mechanical behavior that accompanies it, at high spatial resolution \citep{peaucelle2011arabidopsis,masud2024correlating,hasib2026sewing}.
In primary walls, AFM has resolved individual cellulose microfibrils, their lamellar organization within the matrix, and their movements under applied stretch \citep{zhang2014visualization,zhang2016spatial,zhang2017nanoscale}.
The principal limitation is sampling.
Specimen preparation, scanning, and analysis restrict the number of high-quality observations that can be collected, so a small AFM dataset captures only a limited portion of an intrinsically variable microstructure.
This creates a central challenge for multiscale mechanics: how can a small number of nanoscale observations be used to characterize the range of mechanical responses that may emerge at larger scales?

One route to this question is computational homogenization, which connects nanoscale structure to effective material behavior \citep{zohdi2008introduction,geers2010multi}.
The underlying idea is that a macroscale material (for example, a plant fiber or a stalk) is composed of heterogeneous microstructures.
We select representative microstructures and subject them to prescribed mechanical loading instead of resolving every variation across the full macroscale domain.
Their responses are averaged to obtain effective material properties, which are then upscaled and used to describe the behavior of the larger structure.
In plant materials, multiscale micromechanics models informed by microimaging have predicted the macroscopic stiffness and strength of bamboo culms and oat stems \citep{gangwar2019microimaging,gangwar2021multiscale}.
Such approaches, however, require a structural description that adequately represents the underlying material.
When only a small number of microscopy observations are available, a single observed microstructure may not capture the variability needed for reliable homogenization.
The challenge is therefore to connect scales and to construct a sufficiently rich ensemble of plausible microstructures from limited experimental observations.

Machine learning has broadened what can be done with limited AFM data, including automated experiment control \citep{kalinin2021automated}, cell detection and probe navigation \citep{rade2022deep}, defect classification \citep{biswas2025conversational}, and image segmentation and property prediction \citep{masud2024machine}.
Synthetic data have become an important part of this toolbox.
\citet{millan2026synthetic} developed \textit{SimuScan}, which generates synthetic AFM images with ground-truth masks and AFM-like artifacts for segmentation and targeted imaging, reducing dependence on manually annotated datasets.
\citet{huang2025improving} used unpaired style translation to close the gap between simulated and experimental AFM images.
\citet{pacheco2024deep} synthesized nanoindentation force-distance curves from contact theory to train a neural network for Young's modulus prediction.
These studies establish synthetic data as a practical strategy for addressing scarce AFM observations.
However, they do not address whether generated microstructures preserve the structural features relevant to mechanical response, nor do they propagate image-level variability into distributions of homogenized mechanical properties.
Diffusion-based generation of AFM microstructures also remains unexplored for plant cell walls.

The unresolved challenge is therefore broader than image augmentation.
Structural and compositional variability in a heterogeneous material implies that sparse observations should lead to a distribution of plausible effective properties \citep{torquato2002random,ghanem2003stochastic}.
This distribution should capture variability arising from the observed microstructure while also exposing uncertainty introduced by the assumptions required to translate that microstructure into mechanical properties.
This issue is particularly important in a developing primary wall, where matrix stiffness depends on composition, crosslinking, and hydration and is not directly measurable at the scale of the AFM scan \citep{cosgrove2005growth}.
Mesoscale simulations of epidermal walls further show that tensile response depends on how the cellulose microfibril network and matrix polysaccharides are organized and interact \citep{zhang2021molecular}.
A multiscale framework for sparse AFM data should therefore retain variability across microstructures, propagate uncertainty in constituent properties, and carry both sources of variation into the effective mechanical response.

\begin{figure}[!htbp]
    \centering
    \includegraphics[width=\linewidth]{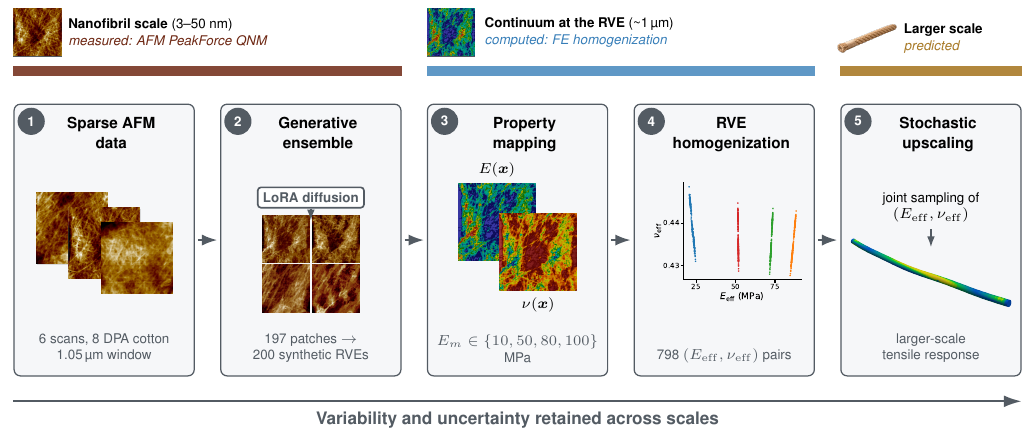}
    \caption{Overview of the generative-physics workflow. Limited AFM scans of 8~DPA cotton fibers (\secref{sec:material}) seed a LoRA-adapted diffusion model that produces an ensemble of candidate synthetic microstructures (\secref{sec:synthetic}). The microstructures are mapped to spatially varying elastic property fields $(E(\bm{x}),\nu(\bm{x}))$ (\secref{sec:mapping}) and homogenized to effective properties $(E_{\mathrm{eff}},\nu_{\mathrm{eff}})$ over a sweep of matrix stiffness values (\secref{sec:homogenization}). The resulting paired-property ensemble is sampled in a larger-scale tensile simulation (\secref{sec:upscaling}). The five stages carry the variability seen in the AFM scans through to the macroscale response. Thumbnails are representative images and results from this study.}%
    \label{fig:multiscalemodeling_process}
\end{figure}

We address this problem with a generative-physics workflow that links sparse AFM observations of 8~DPA (days post-anthesis) cotton fibers to an ensemble of effective mechanical responses (\figref{fig:multiscalemodeling_process}).
Generative modeling first expands the available AFM scans into an ensemble of candidate microstructures \citep{bostanabad2018computational}.
We then map these microstructures to spatially varying elastic fields while systematically varying the uncertain matrix-to-fibril stiffness ratio.
Strain-controlled finite element homogenization converts each mapped microstructure into effective properties $(E_{\mathrm{eff}}, \nu_{\mathrm{eff}})$, producing an ensemble of constitutive pairs.
We finally propagate the paired effective-property ensemble to a larger-scale mechanical simulation.
The workflow is designed to preserve variability across scales. It does not collapse sparse nanoscale observations into a single deterministic material description.
In this work, we make three contributions.
\begin{itemize}
    \item We adapt a latent diffusion model using Low-Rank Adaptation (LoRA), with text and image conditioning, to expand a limited AFM dataset of developing cotton fiber walls into a diverse microstructure ensemble.
    \item We build a high-throughput image-to-property and homogenization pipeline that converts the ensemble into distributions of effective elastic properties and exposes their sensitivity to the uncertain matrix stiffness.
    \item We develop a stochastic upscaling procedure that carries the paired effective properties into a larger-scale simulation while retaining the variability obtained from the microscale ensemble.
\end{itemize}

\section{Materials and Methods}
\label{sec:methods}

\subsection{Plant Material, AFM Imaging, and Nanomechanical Measurements}
\label{sec:material}

We adapted the sample preparation and nanoscale characterization procedures from our previous nano-omics study of developing cotton fibers \citep{hasib2026sewing}.
The present study focuses on \textit{Gossypium hirsutum} (cultivar TM-1) at 8 days post-anthesis (DPA).
Cotton plants were cultivated in a growth chamber (Conviron E-15, Controlled Environments Inc., Pembina, ND, USA) maintained at 28~$^\circ$C. Flowers were self-pollinated, tagged on the day of anthesis, and harvested at 8~DPA.
Bolls were collected at mid-day and stored at $-80$~$^\circ$C prior to experimentation to preserve the native structural integrity of the fibers.
The experimental dataset comprised six raw AFM images obtained across three biological replicates.

For AFM sample preparation, frozen bolls were equilibrated to room temperature before dissection.
The boll wall was removed with fine-tipped tweezers (Cole-Parmer, Straight Fine Tips, Serrated, 101~mm) to isolate individual fibers while minimizing mechanical deformation.
Fibers were incubated in 0.5~M potassium hydroxide (KOH) at room temperature for 30~min to remove loosely bound surface contaminants and non-cellulosic residues, followed by a gentle rinse in deionized water for approximately 3~min to neutralize residual alkali.
Cleaned fibers were mounted on glass microscope slides (Fisherbrand Double Frosted, $25 \times 75 \times 1.0$ mm) and air dried for 10~min at room temperature to ensure stable adhesion during imaging.

AFM experiments were conducted with a BioScope Resolve atomic force microscope (Bruker) operating in PeakForce Quantitative Nanomechanical Mapping (PeakForce QNM) air mode for simultaneous topographical and nanomechanical characterization.
We used a Bruker SNL-10 cantilever with a nominal spring constant of approximately 0.35~N/m and a tip radius of 2~nm.
Deflection sensitivity and spring constant were calibrated before each experiment.
Imaging was performed inside a vibration and acoustic isolation chamber at a scan rate of 0.450~Hz, a PeakForce frequency of 1~kHz, a PeakForce amplitude of 250~nm, and a PeakForce setpoint of approximately 750~pN.
Each raw AFM scan spanned $1.05~\mu\mathrm{m} \times 1.05~\mu\mathrm{m}$.
PNG images exported from NanoScope Analysis and used for downstream image processing were $576 \times 576$ pixels.

We quantified microfibril organization using crossover count and crossover angle.
Crossover count was determined from the number of visible intersections between neighboring microfibrils within selected scan regions.
Crossover angles were measured from the intersection geometry of adjacent microfibrils using NanoScope Analysis software 3.00.
Young's modulus was obtained from force-distance curve analysis with a Hertz model fit \citep{kontomaris2018hertz} at selected points on microfibrils from independent scan regions.
The line-profile data shown in Supplementary Figure S1 contain 40 measurements (17, 12, and 11 points across three images).

\subsection{Matrix-to-Fibril Stiffness Ratio as an Uncertain Model Input}
\label{sec:matrix_stiffness}

The primary cell wall at 8~DPA contains pectin, hemicellulose, and nascent cellulose, and its local elastic response depends on composition, crosslinking, hydration, and structural organization \citep{kim2001cotton,haigler2012cotton}.
Constituent fractions alone therefore do not uniquely determine the matrix elastic constants $(E_m,\nu_m)$.
Direct measurements of the isolated matrix phase are also unavailable at this developmental stage in the present dataset.
The microfibrils, in contrast, have a measured modulus. AFM nanomechanical mapping of 8~DPA \textit{G.~hirsutum} fibers gives a mean microfibril Young's modulus of $E_f=55.1$~MPa \citep{hasib2026sewing}.

We therefore anchor the fibril phase at this measured value and treat the matrix stiffness relative to it, the ratio $E_m/E_f$, as the uncertain model input.
We select four ratios, $E_m/E_f \in \{0.18, 0.91, 1.45, 1.81\}$, corresponding to $E_m \in \{10, 50, 80, 100\}$~MPa.
These ratios span a matrix that is much more compliant than the fibrils, one of nearly equal stiffness, and two that are stiffer than the fibrils.
Sampling on both sides of $E_m/E_f=1$ lets us examine how the effective response changes as the fibrils shift from the stiffer phase to the more compliant phase, without committing to a single unverified matrix modulus.
In all cases the matrix is treated as nearly incompressible with $\nu_m = 0.49$.
Each ratio is propagated through the same homogenization ensemble, so its influence on the resulting effective-property distributions can be evaluated directly, and we use $E_m/E_f$ to label those distributions throughout.

\subsection{Generative Augmentation of the AFM Dataset}
\label{sec:synthetic}

The generative stage expands the limited 8~DPA AFM dataset into a larger ensemble of candidate microstructures for subsequent mechanical analysis.
AFM images can contain artifacts associated with tip condition, multiple contacts, contamination, optical interference, tracking errors, high-frequency ranging, and bow effects \citep{biswas2025conversational}.
The augmentation procedure therefore seeks to preserve domain-specific fibrillar morphology while generating structural variation suitable for ensemble analysis.

We first evaluated the single-image generative adversarial network SinGAN \citep{shaham2019singan}.
The generated images exhibited insufficient morphological diversity, consistent with the model's dependence on patch statistics from a single training image \citep{wang2025singan}.
We therefore use a text-and-image conditioned latent diffusion model \citep{rombach2022high}, which performs denoising in a compressed latent space and supports controllable image synthesis \citep{osorio2024latent}.

Stable Diffusion v1.5 provides a generative prior learned from natural images, whereas AFM micrographs of cellulose contain domain-specific fibrillar morphology and contrast.
We adapt the pretrained model using Low-Rank Adaptation (LoRA) \citep{hu2021lora,wang2024lora}.
LoRA keeps the pretrained base weights frozen and trains low-rank update matrices with rank $r = 4$ and scale $\alpha = 4$ in the cross-attention projections of the U-Net.
The governing equations for latent diffusion, attention, and the LoRA update are provided in Supplementary Section S1.

Before training, we removed bow artifacts and scanning drift using plane-flattening in NanoScope Analysis \citep{bruker_nanoscope_analysis}.
Each of the six $576 \times 576$ exported AFM images was divided into a $3 \times 3$ grid of nine non-overlapping $192 \times 192$ patches, giving 54 original patches.
Each patch was augmented using horizontal flipping, vertical flipping, and combined horizontal and vertical flipping.
Including the original patch, this produced 216 candidate images.
Nineteen images that did not contain microfibrils were excluded, leaving 197 images in the final training dataset.
The retained patches were supplied to Stable Diffusion v1.5 at its $512 \times 512$ input resolution.
Assuming that the exported square spans the full $1.05~\mu\mathrm{m}$ scan without borders or additional cropping, the nominal exported-image spacing is $1050/576 \simeq 1.823$~nm per pixel and each training patch spans $350 \times 350$~nm.
Resizing such a patch to $512 \times 512$ gives a nominal sampling interval of $350/512 \simeq 0.684$~nm per pixel; this resampling does not improve the AFM instrument resolution.
These dimensions describe the training patches and do not establish the physical size of the separately selected homogenization images.
We fine-tuned the LoRA adapter for 12,000 iterations with AdamW, a learning rate of $1 \times 10^{-4}$, cosine annealing, and bf16 mixed precision (Supplementary Table S1).

\begin{figure}[tbp]
    \centering
    \includegraphics[width=.88\linewidth, trim={0.2in 0.1in 0.3in 0.1in},clip]{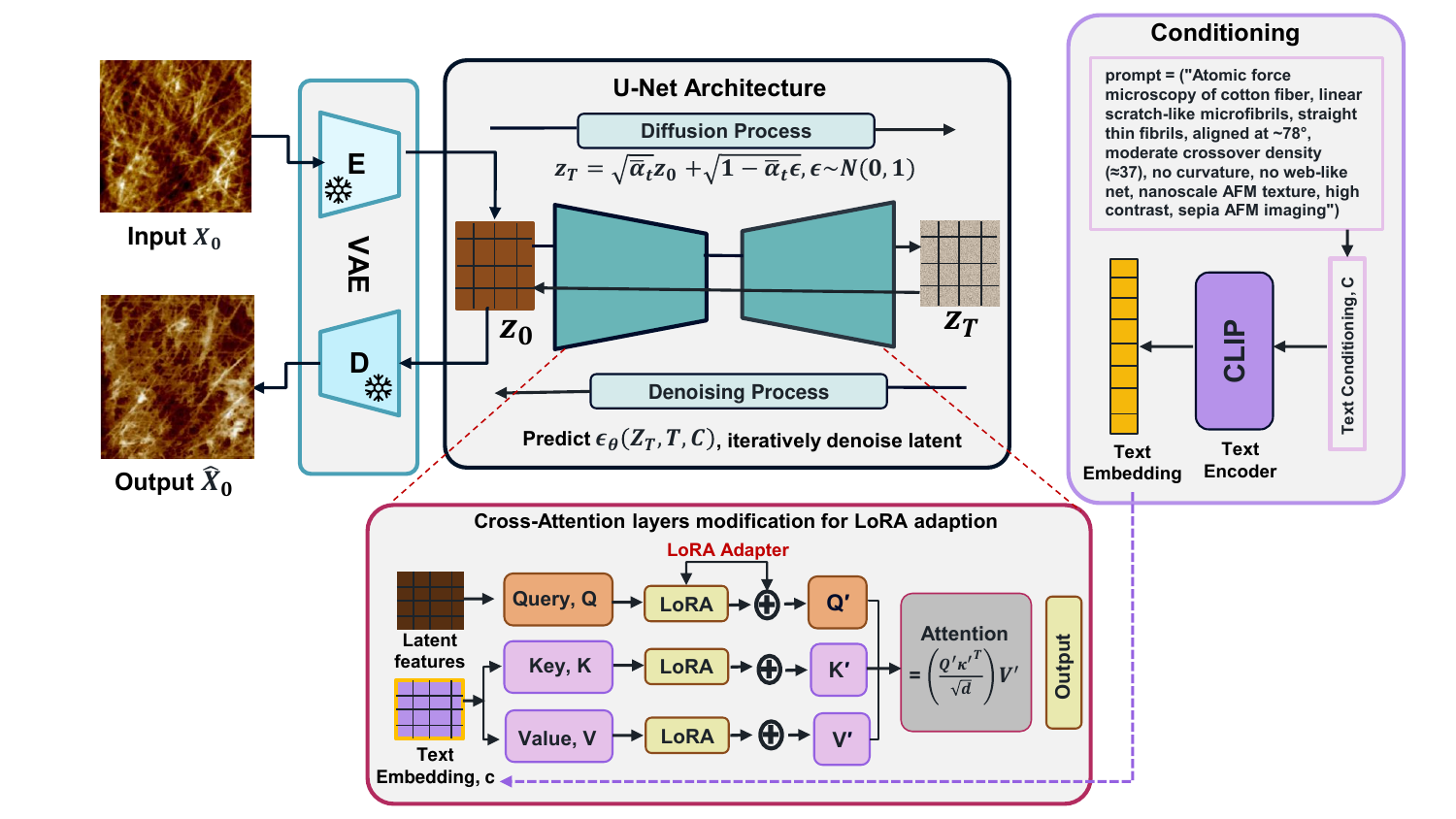}
    \caption{Diffusion pipeline using Stable Diffusion with Low-Rank Adaptation (LoRA) to generate synthetic AFM micrographs of cotton fibers.
        A frozen variational autoencoder (VAE) encoder maps an AFM image patch to a latent representation.
        Noise is progressively added to this latent, and a U-Net denoises it under the embeddings of an input text prompt through LoRA-augmented cross-attention.
        A frozen VAE decoder reconstructs the denoised latent into an image.
        The LoRA weights are the only trained parameters in the pipeline.}
    \label{fig:main_method}
\end{figure}

For inference, the trained LoRA adapter was loaded into the frozen Stable Diffusion backbone.
We conditioned the image-to-image pipeline on a reference experimental AFM image and the text prompt reported in Supplementary Section S1.
The generation used a denoising strength of 0.6, a classifier-free guidance scale of 6 \citep{ho2022classifier}, and 50 inference steps (\figref{fig:main_method}; Supplementary Algorithm S1).
Synthetic variants were generated with different random seeds.
The supplied code draws an integer seed uniformly from 0 to 999999 for each variant.
We treat the resulting images as candidate microstructures, and not as direct replacements for experimental observations, and assess their structural characteristics before using them in the homogenization ensemble.

\subsection{Structural Comparison of Real and Synthetic Images}
\label{sec:crossover_methods}

Because the synthetic images are subsequently used as inputs to the mechanical analysis, we assess whether they preserve selected descriptors of microfibril organization observed in the experimental AFM images.
We use two descriptors that can be measured directly from AFM topography: microfibril crossover count and crossover angle. These descriptors capture structural features that prior work links to the mechanical response of primary walls. Fibril-fibril contacts act as junctions that carry and relax wall stress \citep{park2012revised,zhang2021molecular}, and the net alignment angle of microfibrils sets the anisotropy of wall stiffness and extensibility \citep{baskin2005anisotropic,zhang2021molecular}.

Three real AFM images were used for this comparison.
Eight synthetic variants were evaluated for Image~1, eight for Image~2, and seven for Image~3, giving 23 synthetic images.
Real and LoRA-generated images were analyzed with a semi-automated MATLAB routine that identified and counted microfibril crossover events.
For each identified junction, we measured the acute angle between the two intersecting microfibril trajectories and computed the mean crossover angle for each image.
We report the crossover count per analyzed image. The supplied measurements do not establish whether the validation images are full scans or cropped patches, so an area-normalized density is not reported.

For each real image, the corresponding synthetic descriptor was summarized as the unweighted mean across its variants.
The comparison therefore contains three image-level pairs.
We recomputed two-sided paired $t$-tests on synthetic-minus-real differences, with two degrees of freedom, and report paired standardized effects $d_z=\bar{d}/s_d$.
The 23 synthetic variants are nested within the three conditioning images and are not treated as 23 independent experimental replicates.
Descriptive group means and sample standard deviations give equal weight to the three images.
\footnote{With only three pairs, we note that the tests are exploratory; the distribution of paired differences and independence across biological replicates cannot be established from the supplied image identities.}
Individual measurements and the paired analysis are provided in Supplementary Section S1.

\subsection{Image-to-Property Mapping}
\label{sec:mapping}

To connect image-level structural variability to mechanics, we convert each real or synthetic AFM image into a spatially varying elastic material field.
Normalized image intensity $I(\mathbf{x}) \in [0,1]$ is mapped to local Young's modulus according to
\begin{equation}
E(\mathbf{x}) = E_m + I(\mathbf{x})\left(E_f-E_m\right),
\label{eq:mapping}
\end{equation}
where $E_m$ and $E_f$ denote the matrix-side and fibril-side endpoint values. The fibril-side modulus $E_f=55.1$~MPa is the mean Young's modulus of cellulose microfibrils in 8~DPA \textit{G.~hirsutum} fibers, obtained from Hertz-model fits of PeakForce QNM force curves on 50 randomly selected microfibrils \citep{hasib2026sewing}.
Note that this 50-microfibril dataset from the cited study is separate from the 40 line-profile measurements shown in Supplementary Figure S1.
It is held fixed across all four stiffness-ratio cases of \secref{sec:matrix_stiffness}, so only the matrix-side endpoint $E_m$ changes between cases.
Poisson's ratio is mapped using the same phase endpoints,
\begin{equation}
\nu(\mathbf{x}) = \nu_m + I(\mathbf{x})\left(\nu_f-\nu_m\right),
\qquad \nu_m=0.49.
\label{eq:poisson_mapping}
\end{equation}
We take $\nu_f=0.30$, the value assigned to cellulose microfibrils in homogenization models of plant cell walls \citep{ptashnyk2016impact}, which lies within the range of 0.30 to 0.46 measured for crystalline cellulose by X-ray diffraction \citep{nakamura2004poisson}.
Thus $I=0$ corresponds to the matrix-side endpoint and $I=1$ to the fibril-side endpoint.
When $\nu_f<\nu_m$, Poisson's ratio decreases with intensity.
The endpoint notation does not assume that the fibril-side modulus exceeds the matrix-side modulus in every prescribed stiffness case.

Grayscale intensity serves only as a proxy variable in this mapping.
It does not directly measure local composition or modulus.
This linear intensity-to-modulus assignment follows the practice of image-based finite element modeling, where local image intensity is mapped continuously to local stiffness \citep{gangwar2019microimaging,zannoni1998material}.
Calibrated intensity-to-modulus relations in that literature are typically power-law \citep{helgason2008mathematical}; we adopt linear endpoint interpolation as a simple, uncalibrated form that propagates structural variability into the homogenization.
We use continuous interpolation instead of threshold-based segmentation because Young's modulus measured along line segments in the AFM scans varies continuously, with substantial local fluctuation and no clear binary separation into two phases (Supplementary Figures S1 and S2).
A hard threshold would replace this continuous variation with discrete material regions.
All results reported in the main text therefore use the continuous mapping.

\subsection{Computational Homogenization and Simulation Ensemble}
\label{sec:homogenization}
\label{sec:simulation_workflow}

Each mapped image defines a two-dimensional computational cell with spatially varying $(E(\bm{x}),\nu(\bm{x}))$.
We retain the term representative volume element (RVE) for these computational cells, but statistical representativeness has not been established by a correlation-length or window-size study.
We model each cell as a heterogeneous linear elastic material under an imposed plane-strain assumption and compute its apparent effective response using strain-controlled numerical homogenization.

We impose three linearly independent macroscopic strain states through affine displacement boundary conditions on a unit RVE: unit longitudinal strain, unit transverse strain, and unit engineering shear.
The resulting stress fields are volume averaged for each load case, and the three averaged stress vectors form the columns of the effective stiffness matrix $C_{\mathrm{eff}}$.
In-plane isotropy requires equality of the normal stiffnesses, negligible normal-shear coupling, and the isotropic shear identity. Matrix symmetry alone is insufficient.
We screen every run for approximate isotropy with a 1\% tolerance on reciprocity ($C_{12}$ versus $C_{21}$), on the normal-stiffness mismatch between $C_{11}$ and $C_{22}$, and on the normal-shear coupling terms $C_{13}$ and $C_{23}$ relative to $C_{11}$ (Supplementary Section S3).
The screen does not test the isotropic shear identity, so passing it indicates approximate, not complete, in-plane isotropy.
When the response is treated as approximately isotropic, the effective Young's modulus and Poisson's ratio are obtained from the plane-strain identities $\nu_{\mathrm{eff}} = C_{12}/(C_{11}+C_{12})$ and $E_{\mathrm{eff}} = 2C_{33}(1+\nu_{\mathrm{eff}})$.
The complete homogenization formulation is given in Supplementary Section S3.

Each image is represented on a $512 \times 512$ element grid.
Material properties are evaluated at the finite element quadrature points using $3 \times 3$ Gauss points per element.
The local image intensity at each Gauss point determines the corresponding $E$ and $\nu$ used in the stiffness assembly.
We solve the heterogeneous linear-elastic system for each load case with a direct solver in an octree-based finite element framework \citep{saurabh2021industrial,mehdiOctree}, at a cost of approximately 12 minutes per RVE.

The intended continuous-interpolation ensemble contains 200 microstructures for each of the four prescribed stiffness ratios.
The supplied result files contain 798 completed homogenization runs: 200 for $E_m=10$~MPa, 199 for $E_m=50$~MPa, 199 for $E_m=80$~MPa, and 200 for $E_m=100$~MPa.
The two runs missing from the 800 intended, one each at $E_m=50$ and $E_m=80$~MPa, failed without a stored homogenization summary (Supplementary Table S5).
The result paths are organized under five parent images, with 40 entries per parent in the complete cases.
The resulting database of paired $(E_{\mathrm{eff}},\nu_{\mathrm{eff}})$ values is summarized using Gaussian kernel density estimation with Scott's bandwidth rule \citep{scott1992multivariate,silverman2018density} and subsequently used for stochastic material assignment in the larger-scale simulation.

\subsection{Stochastic Upscaling to a Larger-Scale Simulation}
\label{sec:upscaling}

The final stage tests how the ensemble of homogenized effective properties can be propagated into a larger-scale mechanics calculation without collapsing it to a single constitutive pair.
This stage addresses the question posed in the introduction: how a small number of nanoscale observations can characterize the range of mechanical responses that emerge at larger scales.
The homogenization database contains $N_s$ paired effective properties $\{(E_{\mathrm{eff}}^{(j)}, \nu_{\mathrm{eff}}^{(j)})\}_{j=1}^{N_s}$, which define an empirical joint distribution.
We sample the two properties jointly because $E_{\mathrm{eff}}$ and $\nu_{\mathrm{eff}}$ may be statistically dependent.
Independent sampling from their marginal distributions could produce property combinations that are absent from the homogenization ensemble.

For each Gauss point $\mathbf{x}_g$ of the larger-scale mesh, we draw an index $J_g$ uniformly from the homogenization database and assign the corresponding paired properties,
\begin{equation}
\left( E(\mathbf{x}_g), \nu(\mathbf{x}_g) \right) = \left( E_{\mathrm{eff}}^{(J_g)}, \nu_{\mathrm{eff}}^{(J_g)} \right), \qquad J_g \sim \mathcal{U}\{1,\ldots,N_s\},
\label{eq:joint_empirical_sampling}
\end{equation}
which is equivalent to sampling directly from the empirical joint distribution.
We construct the local constitutive tensor from the sampled pair at each Gauss point and assemble the macroscale linear-elasticity weak form using these Gauss-point-dependent tensors (Supplementary Section S4).

Samples are drawn independently across Gauss points.
The resulting material field therefore preserves the one-point joint distribution of the homogenized properties but does not impose a spatial correlation length.
Spatially correlated sampling can be naturally incorporated once information about the macroscale correlation structure becomes available (and this structure is larger than the distance between Gauss points).

The demonstration uses a non-dimensional, gently curved fiber of circular cross-section with diameter $D=0.03$, spanning $0\le x\le L$ with $L=1$ (centerline length 1.01, $L/D\approx34$).
The centerline bends in the $x$-$y$ plane with a lateral offset of 0.155 between its ends, its inclination to the $x$-axis increasing from about $3^\circ$ to $15^\circ$, and it has a small out-of-plane waviness of about $\pm0.008$.
The specimen is discretized with an adaptive octree mesh of linear (first-order) finite elements, using base refinement level 6 with two additional refinement levels near the STL surface (level 8 at the boundary). The resulting mesh has 365{,}872 elements, and each solve is distributed across 10 MPI ranks.
The end face at $x=0$ is fixed, an axial displacement $u_x=10^{-2}$ is prescribed on the opposite end face, and the lateral surface is traction-free.
Because lengths and displacements share the same unit, the prescribed end displacement corresponds to a nominal axial strain of $10^{-2}$.
Note that the nominal strain is defined on the axial span $L=1$, and the strain along the curved centerline (length 1.01) is about 1\% smaller.
The sampled moduli are in pascals, so the computed stresses are in pascals as well and we report them in MPa.
The stochastic material input is the paired ensemble from the $E_m=50$~MPa ($E_m/E_f=0.91$) microscale case.
Because each macroscale run is computationally expensive, we demonstrate the procedure for this single representative stiffness ratio; it applies unchanged to the other ratios.
To quantify how the sampled material field propagates into macroscale observables, we repeat the simulation for 20 independent realizations.
Geometry, mesh, and loading are identical across realizations; only the per Gauss-point assignment of \eqnref{eq:joint_empirical_sampling} is redrawn, using sampling seeds $12345,\ldots,12364$.
For each realization we record the maximum element-averaged von~Mises stress and summarize its distribution across realizations with a Gaussian KDE (Scott's bandwidth rule).
This completes the workflow, and we now report what each of its stages produces.

\section{Results}
\label{sec:results}

We present the results in the order of the workflow.
We first assess the structural validity of the synthetic images, then examine the mapped property fields and their homogenization.
We next report the effective-property distributions and the extremal microstructures that bound them.
Finally, we report the propagation of the effective-property ensemble to a macroscale specimen.

\subsection{Crossover Count and Crossover Angle in Real and Synthetic Images}
\label{result_SDLoRA}

\begin{figure}[tbp]
    \centering
    \includegraphics[width=\linewidth]{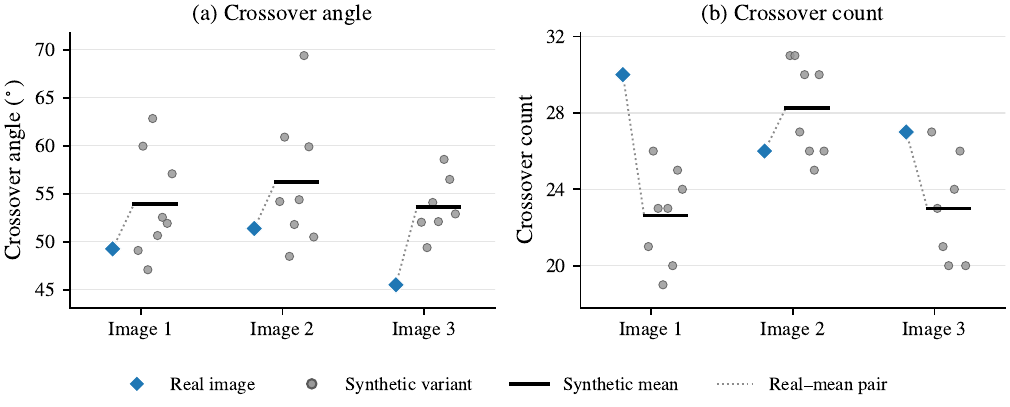}
    \caption{Crossover descriptors for three real AFM images and their image-conditioned synthetic variants.
        (a) Crossover angle and (b) crossover count.
        Blue diamonds denote real-image values, gray circles denote individual synthetic values, black bars denote the synthetic mean for each conditioning image, and dotted lines connect each real value to its synthetic mean.
        Images~1 and~2 each have eight variants and Image~3 has seven; synthetic values are spread horizontally for visibility.
        The paired analysis compares each real value with its corresponding synthetic mean. The data indicates that the generative approach produces synthetic images well aligned with the experimental images.}
    \label{fig:validation_afm_img}
\end{figure}

\figref{fig:validation_afm_img} compares the crossover descriptors measured from the three real images and their LoRA-generated variants.
Real crossover counts were 30, 26, and 27 per analyzed image (mean $\pm$ sample SD: $27.67 \pm 2.08$).
Synthetic per-image means were 22.625, 28.250, and 23.000, based on eight, eight, and seven variants, respectively ($24.63 \pm 3.14$ across the three means).
The mean paired difference (synthetic minus real) was $-3.04$ counts, with a 95\% confidence interval of $[-15.17,9.09]$ counts ($t(2)=-1.08$, $p=0.393$, $d_z=-0.62$).
This small comparison does not detect a count difference, but the wide interval does not establish equivalence.

Real mean crossover angles were $49.27^\circ$, $51.38^\circ$, and $45.52^\circ$ ($48.72^\circ \pm 2.97^\circ$).
The corresponding synthetic per-image means were $53.90^\circ$, $56.19^\circ$, and $53.66^\circ$ ($54.58^\circ \pm 1.40^\circ$).
The mean paired difference was $5.86^\circ$, with a 95\% confidence interval of $[0.95^\circ,10.77^\circ]$ ($t(2)=5.14$, $p=0.036$, $d_z=2.97$).
These exploratory, unadjusted tests use three image-level pairs; the reported group SDs describe variability across image-level means, not across all synthetic variants.

\subsection{Mapped Property Fields and Homogenized Response}
\label{sec:results_mapping}

\begin{figure}[!htbp]
    \centering
    \captionsetup[subfigure]{font=small,labelfont=bf,skip=2pt}
    \begin{subfigure}[t]{0.48\linewidth}
        \centering
        \pvzoompanel{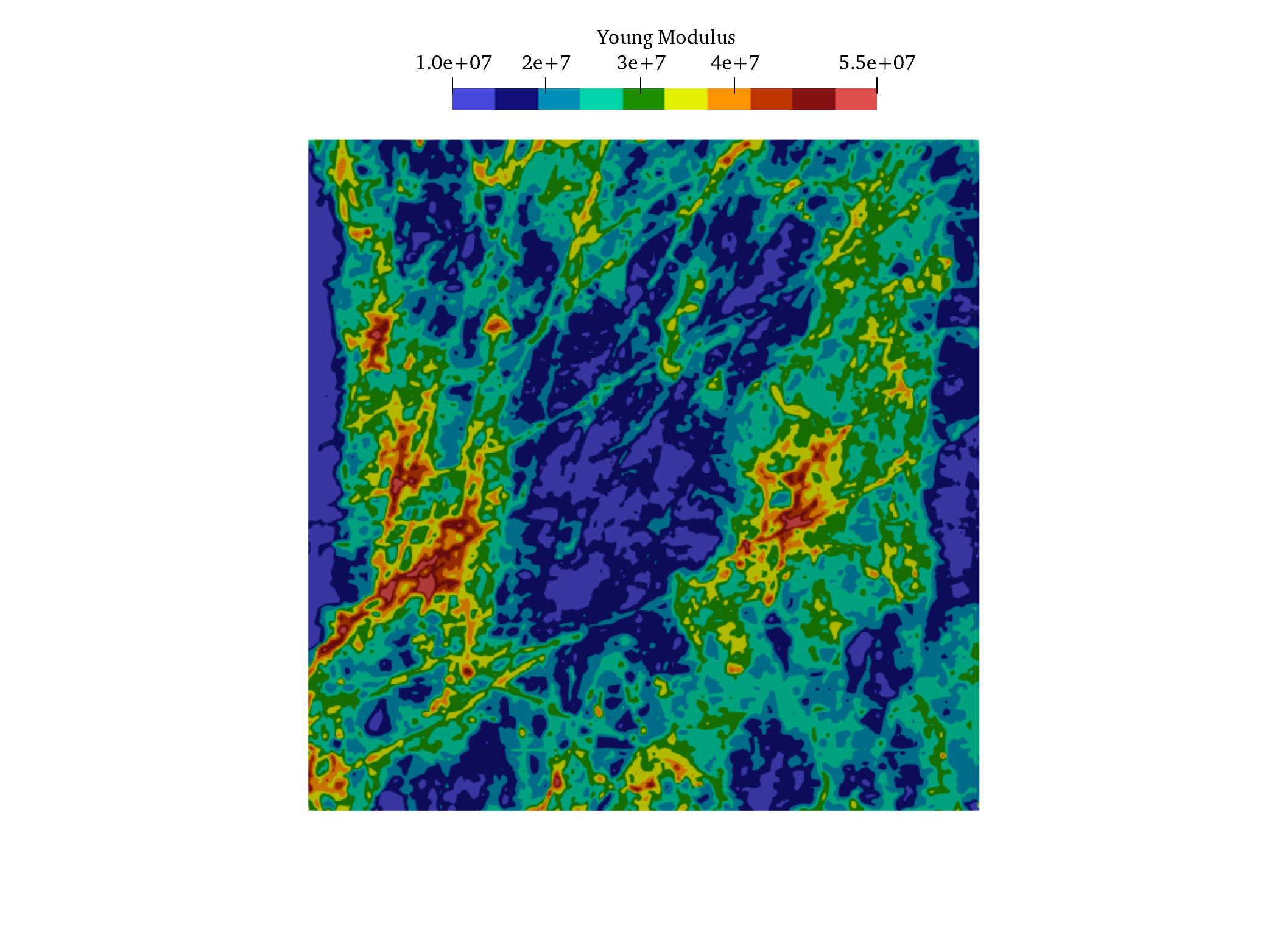}
        \caption{Young's modulus field $E(\bm{x})$.}
        \label{fig:youngs_modulus_field}
    \end{subfigure}
    \hfill
    \begin{subfigure}[t]{0.48\linewidth}
        \centering
        \pvzoompanel{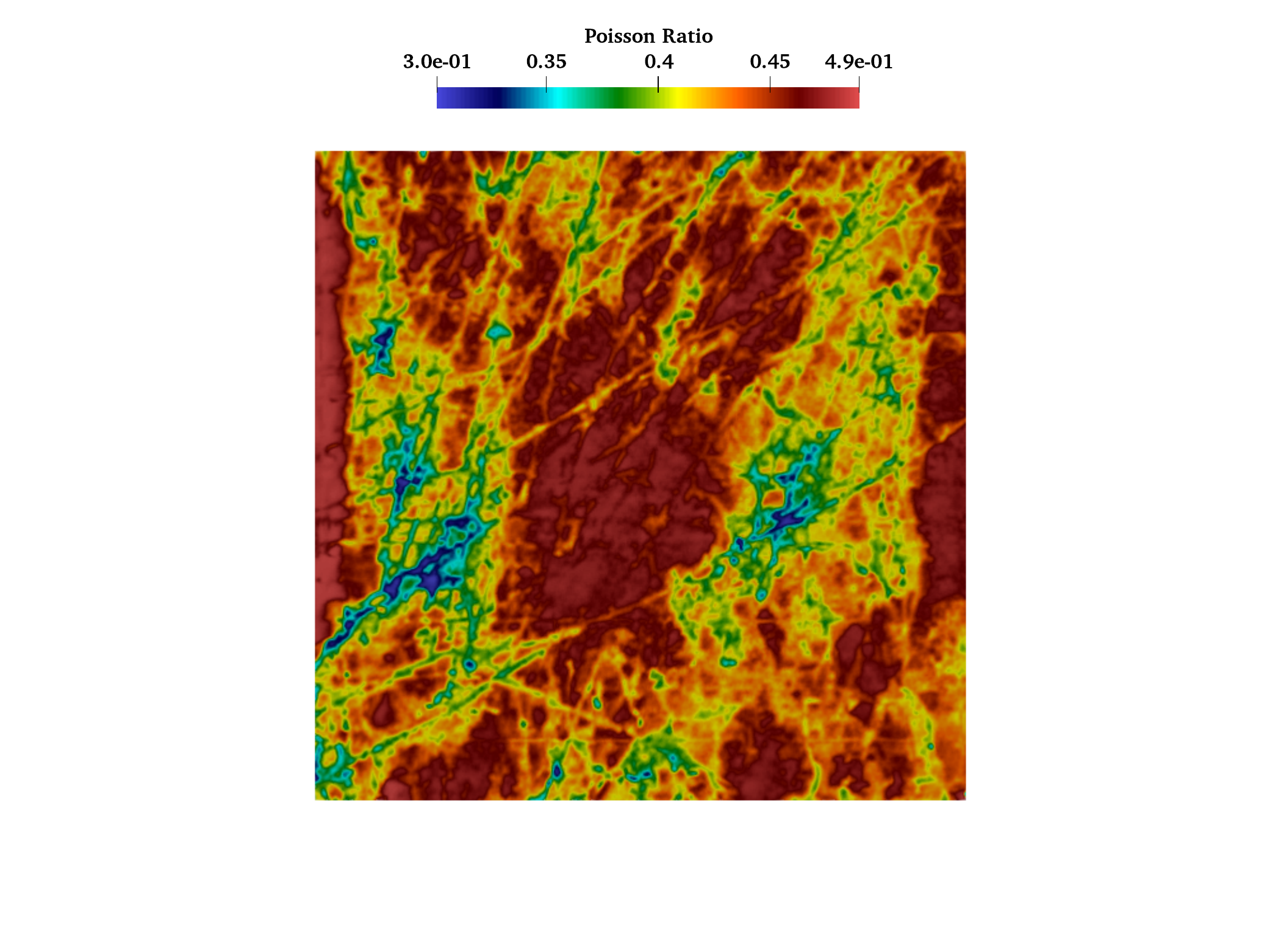}
        \caption{Poisson's ratio field $\nu(\bm{x})$.}
        \label{fig:poissons_ratio_field}
    \end{subfigure}
    \caption{Spatially varying elastic property fields mapped from a representative AFM image using linear intensity interpolation.
        The fields encode spatial variability under the adopted intensity-to-property mapping.
        They are model inputs derived from the normalized AFM intensity proxy.}
    \label{fig:material_props}
\end{figure}

\figref{fig:material_props} shows the Young's modulus and Poisson's ratio fields obtained from a representative AFM image with the mapping of \secref{sec:mapping}.
Grayscale variation produces spatial contrasts in the assigned elastic properties across the RVE.
These fields are model inputs derived from the intensity proxy, and we do not measure the cellulose-rich and pectin-dominated phase composition directly.
Under the three macroscopic load cases, the property heterogeneity drives substantial fluctuations in the microscale displacement fields even though the imposed strain is uniform (Supplementary Figure S3).
The finite element solution resolves these fluctuations, and the fluctuations enter the volume-averaged stresses from which $C_{\mathrm{eff}}$ and the effective constants are extracted.
A mesh-refinement study on one representative image ($E_m=10$~MPa, uniform Q1 meshes from $2^5\times2^5$ to $2^9\times2^9$ elements) shows that these volume-averaged constants are insensitive to resolution (Supplementary Section S5, Table S6 and Figure S5). The $E_{\mathrm{eff}}$ difference from the finest mesh decreases from 1.57\% to 0.076\% between levels 5 and 8, and $\nu_{\mathrm{eff}}$ varies by less than 0.15\% over all levels.

The mesh study and the isotropy screen are the two validity checks that allow the extraction of effective constants from each RVE.
The isotropy screen supports the isotropic reduction used to extract $(E_{\mathrm{eff}},\nu_{\mathrm{eff}})$ for most of the ensemble (Supplementary Table S5).
All 798 completed runs satisfy reciprocity, and every run at $E_m/E_f=0.91$, 1.45, and 1.81 passes all criteria, with normal-stiffness mismatch between $C_{11}$ and $C_{22}$ of at most 0.77\%.
At $E_m/E_f=0.18$, 24 of 200 runs (12\%) exceed the 1\% normal-stiffness tolerance, with a maximum mismatch of 1.30\%; these runs come from three of the five parent images.
Across all ratios, 24 of 798 runs (3.0\%) are flagged, and the median mismatch is 0.33\%.
The mismatch is smallest when the two phases have comparable stiffness and grows with the stiffness contrast, so the scalar constants are least exact for the most compliant matrix.
Because the screen omits the isotropic shear identity, these results establish small directional normal-stiffness mismatch and negligible normal-shear coupling, not full isotropy.

\subsection{Effective Property Distributions Depend Strongly on the Matrix-to-Fibril Stiffness Ratio}
\label{sec:results_moduli}

\begin{figure}[!htbp]
    \centering
    \captionsetup[subfigure]{font=small,labelfont=bf,skip=2pt}
    \distlegend
    \vspace{0.15em}

    \begin{subfigure}[t]{0.48\linewidth}
        \centering
        \distpanel{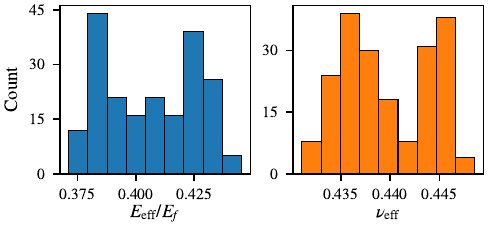}
        \caption{$E_m/E_f = 0.18$}
    \end{subfigure}
    \hfill
    \begin{subfigure}[t]{0.48\linewidth}
        \centering
        \distpanel{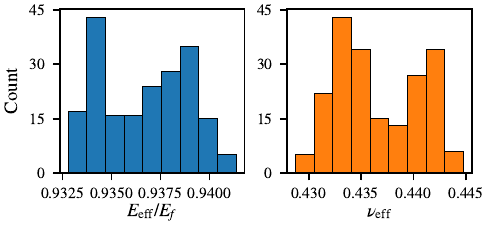}
        \caption{$E_m/E_f = 0.91$}
    \end{subfigure}

    \vspace{0.25em}

    \begin{subfigure}[t]{0.48\linewidth}
        \centering
        \distpanel{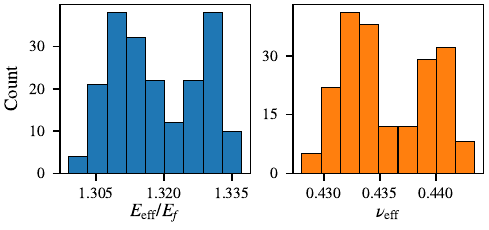}
        \caption{$E_m/E_f = 1.45$}
    \end{subfigure}
    \hfill
    \begin{subfigure}[t]{0.48\linewidth}
        \centering
        \distpanel{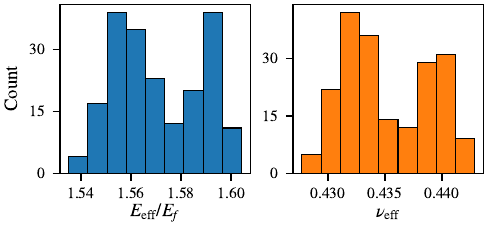}
        \caption{$E_m/E_f = 1.81$}
    \end{subfigure}
    \caption{Distributions of normalized effective Young's modulus $E_{\mathrm{eff}}/E_f$ and Poisson's ratio $\nu_{\mathrm{eff}}$ across the RVE ensemble for matrix-to-fibril stiffness ratios $E_m/E_f=0.18$, 0.91, 1.45, and 1.81 ($E_m=10$, 50, 80, and 100~MPa with $E_f=55.1$~MPa).
        Increasing $E_m/E_f$ shifts the effective-modulus distribution upward, while the relative spread varies nonmonotonically and is smallest for $E_m/E_f=0.91$.}
    \label{fig:moduli_dist}
\end{figure}

\figref{fig:moduli_dist} shows the distributions of $E_{\mathrm{eff}}$ and $\nu_{\mathrm{eff}}$ for each of the four stiffness ratios.
The mean $E_{\mathrm{eff}}$ values are 22.40, 51.61, 72.65, and 86.55~MPa ($E_{\mathrm{eff}}/E_f = 0.41$, 0.94, 1.32, and 1.57) for $E_m/E_f=0.18$, 0.91, 1.45, and 1.81, respectively.
The relative spread is nonmonotonic across the stiffness-ratio sweep: the coefficients of variation of $E_{\mathrm{eff}}$ are 4.95\%, 0.24\%, 0.73\%, and 1.10\%, respectively.
The corresponding mean $\nu_{\mathrm{eff}}$ values are 0.440, 0.437, 0.436, and 0.435, with standard deviations of 0.0045, 0.0040, 0.0039, and 0.0038.
Thus, changing the prescribed stiffness ratio shifts the effective-property distributions and also changes their spread within the adopted image-to-property model.

\begin{figure}[!htbp]
    \centering
    \begin{subfigure}[t]{0.24\linewidth}
        \centering
        \includegraphics[width=\linewidth]{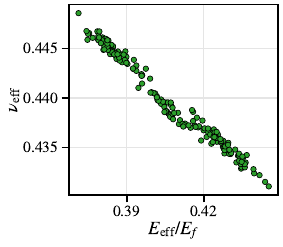}
        \caption{$E_m/E_f = 0.18$}
    \end{subfigure}
    \hfill
    \begin{subfigure}[t]{0.24\linewidth}
        \centering
        \includegraphics[width=\linewidth]{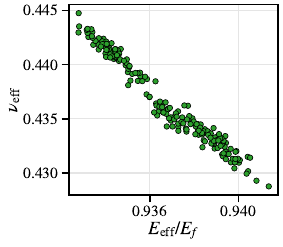}
        \caption{$E_m/E_f = 0.91$}
    \end{subfigure}
    \hfill
    \begin{subfigure}[t]{0.24\linewidth}
        \centering
        \includegraphics[width=\linewidth]{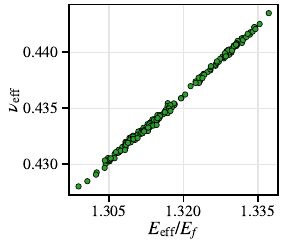}
        \caption{$E_m/E_f = 1.45$}
    \end{subfigure}
    \hfill
    \begin{subfigure}[t]{0.24\linewidth}
        \centering
        \includegraphics[width=\linewidth]{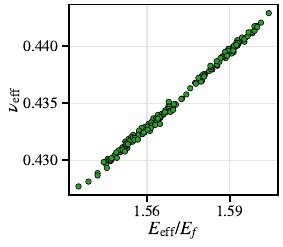}
        \caption{$E_m/E_f = 1.81$}
    \end{subfigure}
    \caption{Normalized effective Young's modulus $E_{\mathrm{eff}}/E_f$ versus effective Poisson's ratio $\nu_{\mathrm{eff}}$ (dimensionless) across the RVE ensemble for each stiffness ratio $E_m/E_f$.
        The sign of the correlation between the two effective constants changes with $E_m/E_f$.
        The correlation is negative for $E_m/E_f = 0.18$ and $0.91$, and positive for $E_m/E_f = 1.45$ and $1.81$.}
    \label{fig:moduli_scatter}
\end{figure}

\figref{fig:moduli_scatter} shows the joint behavior of the two effective constants as a scatter of $E_{\mathrm{eff}}$ against $\nu_{\mathrm{eff}}$.
For $E_m/E_f = 0.18$ and $0.91$ ($E_m = 10$ and $50$~MPa) the relationship is negative, and for $E_m/E_f = 1.45$ and $1.81$ ($E_m = 80$ and $100$~MPa) it is positive.
A first-order explanation follows from the sign of the stiffness contrast. Across microstructures, higher microfibril content moves both local fields toward the fibril-side endpoints. Because $\nu_f < \nu_m$ in all cases, $\nu_{\mathrm{eff}}$ decreases with increasing fibril content.
The effect on $E_{\mathrm{eff}}$ depends on the contrast: for $E_m/E_f < 1$ the fibrils are the stiffer phase and $E_{\mathrm{eff}}$ increases with content, and for $E_m/E_f > 1$ the matrix is stiffer and $E_{\mathrm{eff}}$ decreases.
This predicts a negative joint trend below $E_m/E_f = 1$ and a positive trend above it, consistent with the observed sign change. Confirming that fibril content is the dominant source of variation requires linking each point to quantitative microstructural descriptors, such as the mean image intensity.
This is left as future work.

\begin{figure}[!htbp]
    \centering
    \includegraphics[width=0.95\linewidth]{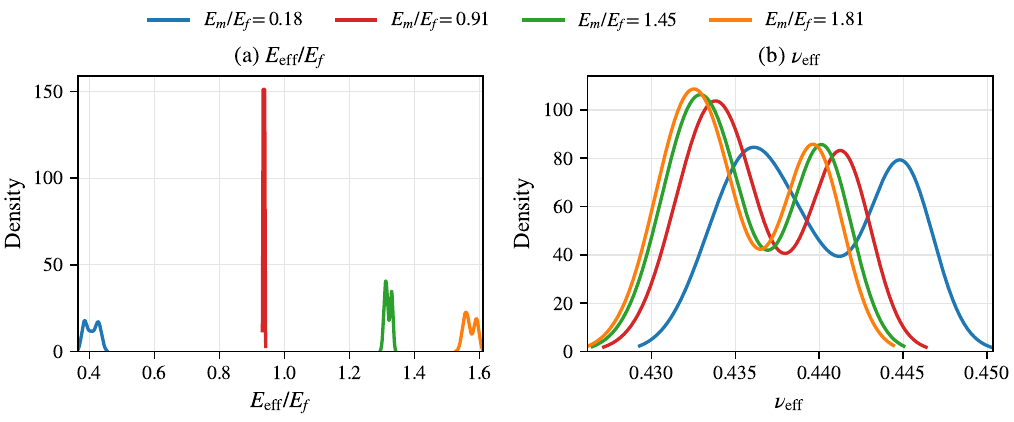}
    \caption{Kernel density estimates of normalized effective Young's modulus $E_{\mathrm{eff}}/E_f$ and Poisson's ratio $\nu_{\mathrm{eff}}$ for $E_m/E_f=0.18$, 0.91, 1.45, and 1.81.
        The modulus densities shift rightward as $E_m/E_f$ increases, while their widths vary nonmonotonically.
        Visible peaks can reflect parent-image grouping and bandwidth choice as well as variation within the sampled ensemble.}
    \label{fig:pdf_all}
\end{figure}

Several of the $E_{\mathrm{eff}}$ distributions appear bimodal, and we summarize the ensemble with kernel density estimates (KDEs).
\figref{fig:pdf_all} overlays the estimated marginal densities of $E_{\mathrm{eff}}$ and $\nu_{\mathrm{eff}}$ for the four stiffness ratios.
The modulus densities shift rightward as $E_m/E_f$ increases, while their widths vary nonmonotonically across the stiffness-ratio sweep.
Several of the curves contain more than one local peak.
The $E_m/E_f = 1.45$ case ($E_m = 80$~MPa), for example, shows two peaks separated by a lower-density region (Supplementary Figure S4).
The multiple peaks describe the sampled ensemble, whose entries are grouped under five parent images.
Parent-image differences, finite sampling, and the KDE bandwidth can all contribute to the apparent modes; repeated variants from a small number of parents do not establish population-level multimodality.
Note that two visible modes do not establish two distinct physical classes of microstructure.
They do motivate sampling from the full empirical distribution in the upscaling step.

\subsection{Extremal Microstructures}
\label{sec:results_extrema}

\begin{figure}[!htbp]
    \centering
    \includegraphics[width=0.32\linewidth]{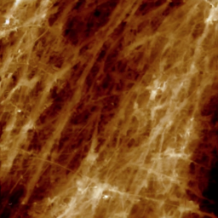}\hfill
    \includegraphics[width=0.32\linewidth]{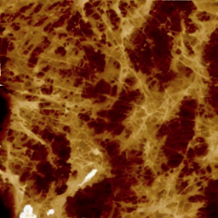}\hfill
    \includegraphics[width=0.32\linewidth]{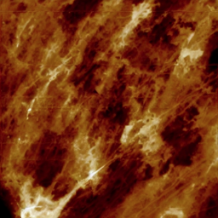}
    \caption{Extremal microstructures from the $E_m/E_f = 1.45$ ($E_m = 80$~MPa) ensemble: the RVE associated with the minimum (left), value closest to the ensemble mean (center), and maximum (right) effective Young's modulus $E_{\mathrm{eff}}$.
        The examples show the spatial organization of the mapped property fields at the extremes and near the mean of the effective-stiffness ensemble.}
    \label{fig:three_extrema}
\end{figure}

We examine the extremal microstructures to show the microstructural spread that underlies the ensemble variability and motivates sampling from the full distribution.
\figref{fig:three_extrema} shows three RVEs from the $E_m/E_f = 1.45$ ($E_m = 80$~MPa) ensemble: the one with the lowest $E_{\mathrm{eff}}$, the one closest to the ensemble mean, and the one with the highest $E_{\mathrm{eff}}$.
The three examples show visible differences in the organization of high- and low-stiffness regions.
These examples illustrate variability within the mapped fields.
Quantitative attribution of the stiffness ranking requires the corresponding image intensities, property endpoints, and effective tensors. Visual differences alone do not establish the cause of the density shape.
This spread is carried into the macroscale response through the stochastic upscaling step reported next.

\subsection{Propagation to a Macroscale Tensile Specimen}
\label{sec:macro_scale_simulation}

\begin{figure}[!htbp]
    \centering
    \captionsetup[subfigure]{font=small,labelfont=bf,skip=2pt}
    \begin{subfigure}[t]{0.48\linewidth}
        \centering
        \macropanel{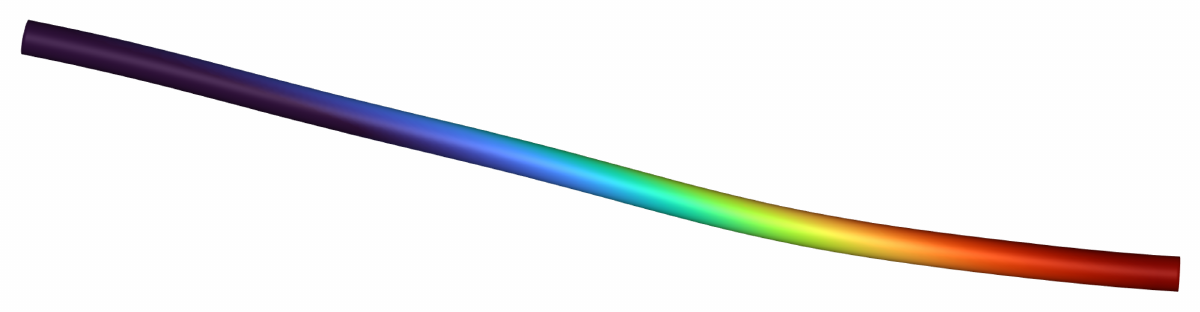}
        \caption{Axial displacement $u_x$.}
    \end{subfigure}
    \hfill
    \begin{subfigure}[t]{0.48\linewidth}
        \centering
        \macropanel{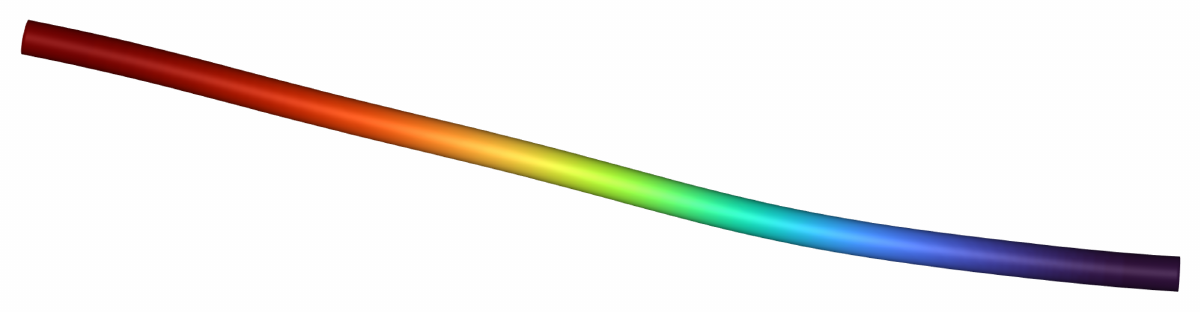}
        \caption{Transverse displacement $u_y$.}
    \end{subfigure}

    \vspace{0.8em}

    \begin{subfigure}[t]{0.48\linewidth}
        \centering
        \macropanel{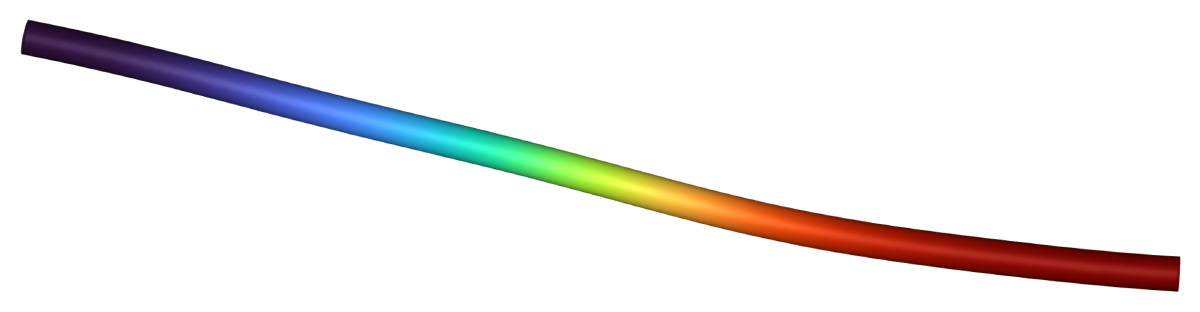}
        \caption{Transverse displacement $u_z$.}
    \end{subfigure}
    \hfill
    \begin{subfigure}[t]{0.48\linewidth}
        \centering
        \macropanel{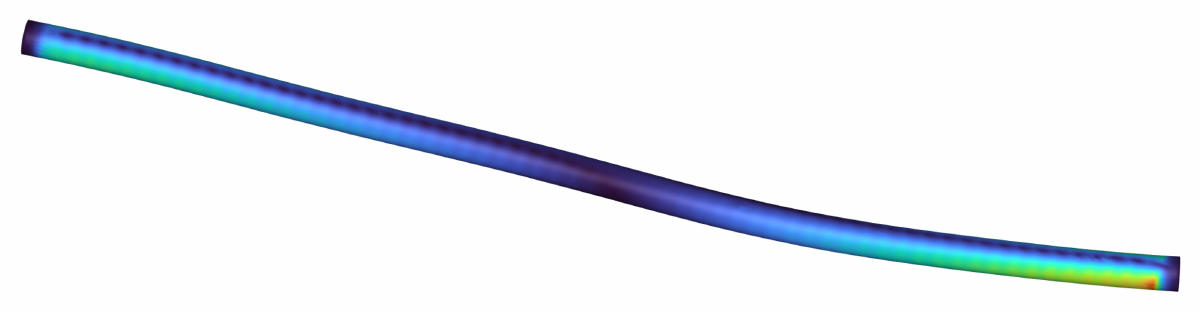}
        \caption{von~Mises stress.}
    \end{subfigure}
    \caption{Macroscale tensile simulation of a non-dimensional curved fiber specimen with axial span $L=1$ and diameter $D=0.03$.
        The end face at $x=0$ is fixed and the opposite end face is prescribed an axial displacement $u_x=10^{-2}$.
        Displacements are non-dimensional (in units of $L$) and the von~Mises stress is in MPa.
        Elastic properties at each quadrature point are sampled from the $E_m/E_f=0.91$ ($E_m=50$~MPa) homogenized effective-property distribution; one of the 20 realizations summarized in \figref{fig:macro_kde} is shown.}
    \label{fig:macro_tension}
\end{figure}

This final stage realizes the third contribution: the stochastic upscaling of the paired effective properties to a larger-scale simulation.
We carry the sparse set of nanoscale observations into a distribution of larger-scale mechanical response while retaining the microscale variability.
We assign a jointly sampled elastic-property pair from the $E_m/E_f=0.91$ ($E_m=50$~MPa) ensemble to each quadrature point of the curved fiber specimen described in \secref{sec:upscaling}.
This creates a heterogeneous constitutive field whose one-point distribution is inherited from the homogenization results.
\figref{fig:macro_tension} shows the displacement components and von~Mises stress for one of the 20 stochastic realizations.
The axial displacement follows the imposed tensile loading.
Because the specimen centerline is curved, axial loading also partially straightens the fiber, so the transverse displacements (up to about $6\times10^{-2}$, several times the imposed axial displacement) are dominated by this geometric effect, and not by lateral contraction.
The resulting bending superposes on the axial tension, and together with the end constraints it produces the spatial gradient in von~Mises stress. This gradient therefore cannot be attributed solely to random material assignment.
Note that this larger-scale calculation is primarily to demonstrate uncertainty propagation. %

\begin{figure}[!htbp]
    \centering
    \includegraphics[width=0.6\linewidth]{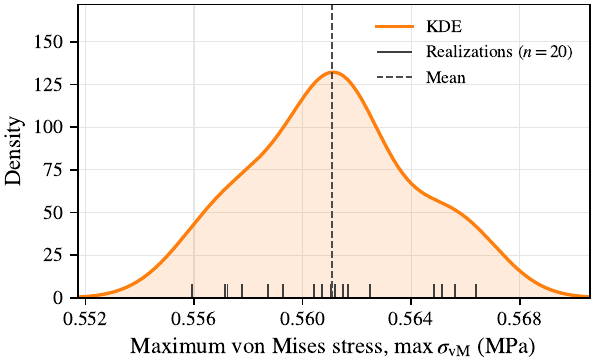}
    \caption{Kernel density estimate of the maximum element-averaged von~Mises stress across 20 macroscale realizations ($E_m/E_f=0.91$, $E_m=50$~MPa).
        Geometry and loading are fixed; only the Gauss-point material assignment is redrawn.
        Vertical ticks mark the individual realizations and the dashed line marks their mean; Gaussian kernel with Scott's bandwidth (0.0016~MPa).}
    \label{fig:macro_kde}
\end{figure}

\figref{fig:macro_kde} summarizes the 20 realizations.
The maximum element-averaged von~Mises stress has a mean of 0.561~MPa and a standard deviation of 0.003~MPa (coefficient of variation 0.52\%), ranging from 0.556 to 0.566~MPa.
For comparison, the nominal axial stress $\bar{E}_{\mathrm{eff}}\,\varepsilon = 51.61~\mathrm{MPa}\times10^{-2} = 0.516$~MPa, so the peak von~Mises stress exceeds it by about 9\%, reflecting curvature-induced bending, the end constraints, and the local heterogeneity.
The maximum axial stress magnitude $\max|\sigma_{xx}|$ (mean 0.81~MPa) varies more, with a coefficient of variation of 3.3\%.
The maximum displacement magnitude is essentially unchanged across realizations (coefficient of variation below $10^{-3}\%$) because it is governed by the prescribed kinematics.
The microscale variability therefore reaches the macroscale observables, but in attenuated form: with properties drawn independently at each Gauss point, local fluctuations average out over each element and over the specimen cross-section.
The narrow spread is thus a lower bound conditional on the absence of spatial correlation; a correlated material field with a correlation length comparable to the specimen diameter would average less and widen these distributions.
Together with the homogenization ensemble, these realizations close the loop from AFM-derived microstructures to a distribution of macroscale structural responses.

\section{Discussion}
\label{sec:discussion}

Sparse AFM observations of a developing cell wall yield a distribution of effective properties whose shape and joint structure depend strongly on the matrix-to-fibril stiffness ratio, and a single nominal constitutive pair does not encode this information.
The workflow presented here produces that distribution and carries it into a larger-scale mechanics calculation.
The generative model expands the set of candidate microstructures, the image-to-property map converts those structures into heterogeneous elastic fields, and ensemble homogenization produces paired effective properties.
The homogenization results show that the effective-property distributions shift with the prescribed matrix-to-fibril stiffness ratio, and the joint relationship between $E_{\mathrm{eff}}$ and $\nu_{\mathrm{eff}}$ changes sign across the stiffness-ratio cases.
Several estimated densities also appear multimodal.

For developing cotton fibers at 8~DPA, local wall mechanics may vary across the observed microstructure, and a multiscale model should retain that variability when the data support it.
The same workflow applies to AFM scans from other developmental stages, genotypes, or growth conditions.
It offers a route for asking how changes in wall organization during fiber development translate into changes in the distribution of effective wall mechanics.

Several limitations currently bound the quantitative use of these results.
One limitation of this study is the generative stage, which is bounded by the experimental images used for training and validation.
The paired comparison does not detect a crossover-count difference, but its wide confidence interval does not establish agreement.
The crossover-angle comparison shows an exploratory difference at the unadjusted 0.05 level.
The 197 training images derive from only six raw scans through patch extraction and flipping; they do not represent 197 independent experimental observations.
The training membership and biological-replicate identities of the three validation images are unreported, so this comparison is not established as an independent held-out test.
Full biological and distributional fidelity remains to be established.
Additional descriptors (for example microfibril width, spacing, and orientation distributions), a larger independent experimental validation set, and tests that compare real and synthetic distributions would strengthen the use of the generative ensemble as a quantitative surrogate for experimental sampling.

A second limitation is the image-to-property mapping, in which normalized AFM grayscale intensity serves as a proxy for spatially varying elastic properties and gives heterogeneous fields that are useful for testing the multiscale pipeline.
Intensity is not itself a composition or modulus measurement.
Co-registered PeakForce QNM modulus and topography data, if retained for the same fields, would provide a starting point for calibrating the mapping, alongside independent composition measurements.
Until such a calibration is performed, the mapped fields should not be read as quantitatively resolved constituent mechanics.

A third limitation is the set of mechanical idealizations used in the homogenization.
The reduction to $(E_{\mathrm{eff}},\nu_{\mathrm{eff}})$ assumes isotropy, although the AFM microstructures contain oriented fibrillar features.
The isotropy screen finds normal-stiffness mismatch below 1.3\% in every run and flags only 3\% of runs, all at the most compliant matrix, but it does not test the isotropic shear identity, and the complete effective tensors were not stored.
Plane strain is an imposed modeling assumption; the surface images alone do not establish the out-of-plane constraint.
The mesh study in Supplementary Section S5 covers a single image and stiffness ratio; correlation-length, window-size, boundary-condition, and locking checks, and mesh checks across the ensemble, are also needed to establish the reliability of the apparent effective properties.
Storing the complete $C_{\mathrm{eff}}$ for every run, testing the shear identity, and retaining an orthotropic effective tensor where isotropy is not supported would make the mechanical interpretation more defensible, particularly at low $E_m/E_f$.
The apparent modes in the effective-modulus densities require a bandwidth-sensitivity analysis and quantitative links between structural descriptors and mode membership before they can be read as distinct physical microstructure classes.

Finally, the homogenization uses two-dimensional microstructures and linear elasticity, whereas a cotton fiber wall is three-dimensional and may exhibit nonlinear, viscoelastic, poroelastic, or moisture-dependent behavior.
We assign the macroscale material samples independently at Gauss points.
The larger-scale field thus preserves the one-point joint distribution without a measured spatial correlation length, and the narrow spread of the 20 macroscale realizations reflects this independent sampling; 20 realizations also limit the resolution of the tails of the response distribution.

The next calibration steps are to establish whether co-registered modulus data are available and to expand the independent structural validation of the synthetic images.
A complete anisotropy assessment that includes the shear identity and a robustness analysis of the apparent multimodality would complete this set.
We anticipate that these steps will turn the workflow into a quantitatively predictive material model for developing plant cell walls.

\section{Conclusions}
\label{sec:conclusions}

We presented a generative-physics workflow that links sparse AFM observations of 8~DPA cotton fiber cell walls to an ensemble of effective elastic properties and propagates that ensemble to a larger-scale mechanics calculation.
The workflow combines LoRA-adapted image generation, an explicit image-to-property mapping, finite element homogenization, empirical probability estimation, and stochastic upscaling.
The homogenization study shows that $E_{\mathrm{eff}}$ and $\nu_{\mathrm{eff}}$ vary across the microstructure ensemble and depend strongly on the prescribed matrix-to-fibril stiffness ratio.
The tensile example shows that these paired properties can be sampled directly in a continuum simulation, and its 20 realizations give a distribution of the macroscale peak stress whose narrow spread reflects spatially independent material sampling.
Quantitative predictive use will require experimental calibration of the intensity-to-property mapping and a larger structural validation of the synthetic images, among the directions detailed in the discussion.
With these limitations made explicit, the paper provides a computational template for uncertainty-aware multiscale characterization of plant cell walls from limited microscopy data.
It offers one route to the question posed at the outset, namely how a small number of nanoscale observations can characterize the range of mechanical responses that emerge at larger scales.
We believe that this template will help connect measurements of wall organization during fiber development to the distribution of effective wall mechanics.

\section*{Data Availability Statement}
The AFM images, synthetic image ensemble, LoRA adapter weights, homogenization outputs, and analysis code supporting the conclusions of this article will be made available by the authors upon reasonable request.

\section*{Author Contributions}
SK: computational homogenization, upscaling simulations, analysis, and writing.
AB: generative model development, training, and writing.
NM and MHHH: cotton fiber preparation, AFM imaging, nanomechanical measurements, and crossover analysis.
AS and BG: conceptualization, supervision, funding acquisition, and writing.
All authors reviewed and approved the final manuscript.

\section*{Funding}
This work was partially supported by the AI Institute for Resilient Agriculture (USDA-NIFA 2021-67021-35329), NSF 2412929, Iowa State University's Plant Science Institute, and DOE GENESIS Phase I project.

\section*{Acknowledgments}
Computing support from ISU HPC and TACC Frontera through NAIRR is acknowledged.
 
\section*{Conflict of Interest}
The authors declare that the research was conducted in the absence of any commercial or financial relationships that could be construed as a potential conflict of interest.

\section*{Supplementary Material}
The Supplementary Material for this article contains five sections.
Section S1 collects the generative model equations, hyperparameters, algorithm, individual crossover measurements, and paired statistical analysis.
Section S2 reports the AFM line-profile measurements that motivate the continuous mapping.
Section S3 gives the full homogenization formulation and the isotropy screen of the effective stiffness tensors, Section S4 the macroscale assembly, and Section S5 additional results, including the displacement fields of the homogenization load cases, the single-stiffness effective-modulus density, and the mesh-convergence study.

\bibliographystyle{elsarticle-harv}
\bibliography{Final}

\end{document}


\section{Generative Model Details}
\label{sec:s_generative}

\subsection{Single-Image GAN Baseline}

SinGAN \citep{shaham2019singan} learns a coarse-to-fine hierarchy of conditional distributions from a single image,
\begin{equation}
p(x_0) = p(x_N)\prod_{n=0}^{N-1} p(x_n \mid x_{n+1}),
\label{eq:singan_hierarchical}
\end{equation}
in which each conditional $p(x_n \mid x_{n+1})$ generates a finer-scale image from the preceding coarser scale.
Its outputs showed insufficient morphological diversity for the present application.
The model is constrained by patch-level statistics from a single training image \citep{wang2025singan}.

\subsection{Latent Diffusion}

Latent diffusion models \citep{rombach2022high} move the diffusion process from pixel space to a compressed latent representation.
A pretrained variational autoencoder (VAE) maps an input image $x$ to $z = \mathcal{E}(x)$, where $\mathcal{E}$ and $\mathcal{D}$ denote the encoder and decoder.
The forward process \citep{Ho2020ddpm} adds noise to the latent representation as
\begin{equation}
z_t = \sqrt{\bar{\alpha}_t}\, z_0 + \sqrt{1-\bar{\alpha}_t}\,\epsilon, \quad \epsilon \sim \mathcal{N}(0,I),
\label{eq:forward_diffusion}
\end{equation}
and the denoising network learns the reverse process by predicting the noise at each timestep with the training objective
\begin{equation}
\mathcal{L} = \mathbb{E}_{z,\epsilon,t}\left[\|\epsilon - \epsilon_\theta(z_t,t,c)\|^2\right],
\label{eq:ldm_loss}
\end{equation}
where $\epsilon_\theta$ is the denoising U-Net and $c$ is the conditioning signal.
Stable Diffusion \citep{rombach2022high,oorloff2025stable} combines a VAE, a latent-space denoising U-Net, and a text encoder based on CLIP \citep{radford2021learning}.
The CLIP encoder maps the text prompt to an embedding that conditions the U-Net through cross-attention using the scaled dot-product attention of \citet{vaswani2017attention},
\begin{equation}
\mathrm{Attention}(Q,K,V)=\mathrm{softmax}\left(\frac{QK^\top}{\sqrt{d}}\right)V,
\label{eq:attention}
\end{equation}
where $Q$, $K$, and $V$ are query, key, and value projections.
Self-attention draws these projections from the same sequence.
Cross-attention uses latent image features as queries and text embeddings as keys and values.
Multi-head attention evaluates these relations across multiple learned subspaces.

\subsection{Low-Rank Adaptation}

For a pretrained attention weight matrix $W_0$, LoRA \citep{hu2021lora,wang2024lora} writes the adapted weight as
\begin{equation}
W' = W_0 + \frac{\alpha}{r}BA,
\label{eq:lora_update}
\end{equation}
and the forward mapping becomes
\begin{equation}
h = W_0x + \frac{\alpha}{r}BAx,
\label{eq:lora_forward}
\end{equation}
where $B \in \mathbb{R}^{d \times r}$ and $A \in \mathbb{R}^{r \times k}$.
The rank satisfies $r \ll \min(d,k)$ and sets the capacity of the adaptation, $x$ is the input feature representation, and $\alpha$ regulates the adaptation strength.
We train only $A$ and $B$ and keep the base Stable Diffusion parameters frozen.
We apply the update to the U-Net cross-attention projections (\texttt{to\_k}, \texttt{to\_q}, \texttt{to\_v}, \texttt{to\_out.0}).

\subsection{Data Preparation, Hyperparameters, and Algorithm}

The AFM data preparation is described in Section~2.3 of the main text.
In brief, six exported AFM images ($576 \times 576$ pixels, $1.05~\mu\mathrm{m} \times 1.05~\mu\mathrm{m}$ fields, three biological replicates) were each split into a $3 \times 3$ grid of $192 \times 192$ patches, augmented with horizontal, vertical, and combined flips, and filtered to 197 training images containing microfibrils.
\tabref{tab:lora_hyperparams} lists the training and inference settings, and \algoref{alg:full_pipeline} gives the corresponding computational workflow.

\begin{table}[ht]
\centering
\caption{Hyperparameter configuration for LoRA fine-tuning and image-to-image inference with Stable Diffusion v1.5 on AFM cotton fiber imagery.}
\label{tab:lora_hyperparams}
\small
\setlength{\tabcolsep}{3pt}
\begin{tabular}{@{}l l l p{0.32\linewidth}@{}}
\toprule
\textbf{Category} & \textbf{Parameter} & \textbf{Value} & \textbf{Effect / Justification} \\
\midrule
\multirow{3}{*}{Base Model}
  & Architecture        & Stable Diffusion v1.5                        & Latent diffusion backbone \\
  & Input resolution    & $512 \times 512$ px                          & Input size used for Stable Diffusion v1.5 \\
  & Precision           & \texttt{bf16}                                & Reduced memory; numerically stable gradients \\
\midrule
\multirow{3}{*}{LoRA Adapter}
  & Rank ($r$)          & 4                                            & Low-rank adaptation \\
  & Alpha ($\alpha$)    & 4                                            & Scaling ratio $\alpha/r = 1.0$ \\
  & Target layers       & \texttt{to\_k, to\_q, to\_v, to\_out.0}      & U-Net cross-attention projections \\
\midrule
\multirow{6}{*}{Training}
  & Optimizer           & AdamW                                        & Weight-decoupled adaptive gradient descent \\
  & $\beta_1,\, \beta_2$& 0.9,\; 0.999                                 & Momentum and variance estimates \\
  & Weight decay        & 0.01                                         & L2 regularization \\
  & Learning rate       & $1 \times 10^{-4}$                           & LoRA fine-tuning learning rate \\
  & LR schedule         & Cosine decay                                 & Learning-rate schedule \\
  & Gradient clip       & 1.0                                          & Gradient clipping \\
\midrule
\multirow{8}{*}{Data}
  & Raw AFM images      & 6                                            & Three biological replicates \\
  & Exported image size & $576 \times 576$ px                           & NanoScope Analysis PNG output \\
  & Field size          & $1.05~\mu\mathrm{m}\times1.05~\mu\mathrm{m}$           & Same physical field size for all raw images \\
  & Patch grid          & $3 \times 3$ per image                       & Nine non-overlapping patches per raw image \\
  & Patch size          & $192 \times 192$ px                          & 54 original patches across six images \\
  & Augmentations       & H, V, and HV flips                           & Original plus three flipped variants \\
  & Candidate images    & 216                                          & 54 patches $\times$ 4 versions \\
  & Final images        & 197                                          & 19 images without microfibrils excluded \\
\midrule
\multirow{2}{*}{Optimization}
  & Effective batch     & 4 (batch 1 $\times$ grad.\ accum.\ 4)        & Effective batch size used in training \\
  & Total steps         & 12{,}000                                     & Fine-tuning iterations \\
\midrule
\multirow{4}{*}{Inference}
  & Pipeline            & \texttt{StableDiffusionImg2ImgPipeline}      & Image and text conditioning \\
  & Denoising strength  & 0.6                                          & Image-to-image denoising strength \\
  & Guidance scale      & 6                                            & Classifier-free guidance scale \\
  & Inference steps     & 50                                           & Denoising steps \\
\midrule
\multirow{2}{*}{Resources}
  & Training time       & $\sim$10 hours                               & Single GPU run (bf16 mixed precision) \\
  & Seed                & Random per generated variant                 & Different seeds generate image variants \\
\bottomrule
\end{tabular}
\end{table}

\subsubsection*{Generation prompt and inference settings}

The image-to-image generation used the following positive prompt:
\begin{quote}
\small
``Atomic force microscopy of cotton fiber, linear scratch-like microfibrils, straight thin fibrils, aligned at $\sim78^\circ$, moderate crossover density ($\approx37$), no curvature, no web-like net, nanoscale AFM texture, high contrast, sepia AFM imaging''
\end{quote}
and the following negative prompt:
\begin{quote}
\small
``curved microfibrils, wavy microfibrils, tangled net, circular blobs, thick fibers, clumps, noisy background, blurry, distorted''
\end{quote}
Generation used a denoising strength of 0.6, a classifier-free guidance scale of 6 \citep{ho2022classifier}, and 50 inference steps.
Different CUDA random seeds were used to generate multiple variants.
The supplied generation code draws the seed as an integer between 0 and 999999 using \texttt{random.randint(0,999999)} and initializes a CUDA generator with that value.
The example loop has three iterations; it is not a record of the complete generation batches, and per-image seeds for the validation and mechanics ensembles were not supplied.

\begin{algorithm}[ht]
\caption{Synthetic AFM Cotton Fiber Image Generation via LoRA-Adapted Stable Diffusion}
\label{alg:full_pipeline}
\begin{algorithmic}[1]

\Require
six exported AFM images $\mathcal{I}$ of size $576\times576$;
SD v1.5 $(\epsilon_\theta,\mathcal{E},\mathcal{D},\tau_\phi)$;
LoRA rank $r=4$, scale $\alpha=4$; target projections;
training settings from Table~\ref{tab:lora_hyperparams};
conditioning image $x_{\mathrm{init}}$; positive/negative prompts $c_+,c_-$
\Ensure synthetic image $\hat{x}$
\State \textbf{Data Preparation}
\State Initialize patch collection $\mathcal{P}\leftarrow\varnothing$
\For{each $I_k\in\mathcal{I}$}
    \State Split $I_k$ into nine non-overlapping $192\times192$ patches
    \State Append these patches to $\mathcal{P}$
\EndFor
\State $\mathcal{P}_{\mathrm{aug}}\leftarrow\{P,\textsc{HFlip}(P),\textsc{VFlip}(P),\textsc{HVFlip}(P)\mid P\in\mathcal{P}\}$
\State Remove 19 images without microfibrils from the 216 candidates
\State $\mathcal{P}_{\mathrm{keep}}\leftarrow$ the 197 retained images
\State $\mathcal{T}\leftarrow\{(\textsc{Resize}_{512}(P),\textsc{Caption}(P))\mid P\in\mathcal{P}_{\mathrm{keep}}\}$
\State \textbf{LoRA Training}
\State Freeze the base U-Net, VAE, and text-encoder weights
\State Initialize trainable LoRA matrices $A,B$ with the training implementation
\State Minimize Eq.~\ref{eq:ldm_loss} on $\mathcal{T}$ using AdamW and the LoRA updates
\State Retain the reported 12{,}000-iteration budget; optimizer-update convention requires confirmation
\State \textbf{Inference (Img2Img)}
\State Load trained LoRA updates $W'=W_0+(\alpha/r)BA$ into the target projections
\State $x_{\mathrm{init}}\leftarrow\textsc{Resize}_{512}(x_{\mathrm{init}})$
\State Draw seed $q\leftarrow\textsc{RandomInteger}(0,999999)$
\State Initialize the CUDA random generator with $q$
\State $\hat{x}\leftarrow\textsc{Img2Img}(x_{\mathrm{init}},c_+,c_-,s=0.6,w=6,T_{\mathrm{inf}}=50,q)$
\State \Return $\hat{x}$

\end{algorithmic}
\end{algorithm}

\subsection{Crossover Validation Measurements}
\label{sec:s_crossover}

Three real AFM images were compared with image-conditioned synthetic variants.
Images~1 and~2 each had eight variants, while Image~3 had seven, for 23 synthetic images.
Tables~\ref{tab:crossover_angle_values} and~\ref{tab:crossover_count_values} reproduce the supplied image-level measurements.
Each angle entry is the mean intersection angle measured within that image.
Counts are per analyzed image, not counts per square micrometer; the crop area for these specific validation images was not supplied.
Synthetic means weight the variants equally within each conditioning image.

\begin{table}[htbp]
\centering
\caption{Mean crossover angle in degrees for each real image and synthetic variant.
    Values are displayed to four decimal places; calculations use the full supplied precision.
    A dash denotes the absent eighth variant for Image~3.}
\label{tab:crossover_angle_values}
\small
\begin{tabular}{@{}lrrr@{}}
\toprule
Measurement & Image~1 & Image~2 & Image~3 \\
\midrule
Real image & 49.2667 & 51.3846 & 45.5185 \\
Variant 1 & 49.0952 & 54.1935 & 52.0370 \\
Variant 2 & 59.9615 & 60.9032 & 49.3913 \\
Variant 3 & 47.0870 & 48.4815 & 54.0952 \\
Variant 4 & 62.8421 & 51.8000 & 52.1000 \\
Variant 5 & 50.6522 & 54.3846 & 58.5833 \\
Variant 6 & 52.5500 & 69.4000 & 56.5000 \\
Variant 7 & 51.9200 & 59.8966 & 52.9000 \\
Variant 8 & 57.0833 & 50.5000 & -- \\
\midrule
Synthetic mean & 53.8989 & 56.1949 & 53.6581 \\
\bottomrule
\end{tabular}
\end{table}

\begin{table}[htbp]
\centering
\caption{Crossover counts for each real image and synthetic variant. A dash denotes the absent eighth variant for Image~3.}
\label{tab:crossover_count_values}
\small
\begin{tabular}{@{}lrrr@{}}
\toprule
Measurement & Image~1 & Image~2 & Image~3 \\
\midrule
Real image & 30 & 26 & 27 \\
Variant 1 & 21 & 31 & 27 \\
Variant 2 & 26 & 31 & 23 \\
Variant 3 & 23 & 27 & 21 \\
Variant 4 & 19 & 30 & 20 \\
Variant 5 & 23 & 26 & 24 \\
Variant 6 & 20 & 25 & 26 \\
Variant 7 & 25 & 30 & 20 \\
Variant 8 & 24 & 26 & -- \\
\midrule
Synthetic mean & 22.625 & 28.250 & 23.000 \\
\bottomrule
\end{tabular}
\end{table}

\subsubsection*{Paired image-level analysis}

Let $r_j$ be the real-image descriptor and $s_{jk}$ the corresponding descriptor for variant $k$ of image $j$, with $n_j=8,8,7$.
We define
\begin{equation}
\bar{s}_j=\frac{1}{n_j}\sum_{k=1}^{n_j}s_{jk},\qquad
 d_j=\bar{s}_j-r_j,\qquad
 t=\frac{\bar{d}}{s_d/\sqrt{3}},\qquad d_z=\frac{\bar{d}}{s_d},
\label{eq:paired_validation}
\end{equation}
where $s_d$ is the sample standard deviation of the three paired differences.
Table~\ref{tab:crossover_paired} reports two-sided paired $t$-tests with two degrees of freedom and 95\% confidence intervals $\bar{d}\pm t_{0.975,2}s_d/\sqrt{3}$.
These are a reanalysis of the supplied measurements.
The tests are exploratory and unadjusted for the two descriptors; neither the normality of the differences nor the biological independence of the three conditioning images can be assessed from the supplied data.
In particular, a nonsignificant count test is not evidence of equivalence.

\begin{table}[htbp]
\centering
\caption{Paired comparison of the synthetic per-image mean with its conditioning real image. Differences are synthetic minus real. Both tests have $n=3$ image pairs and two degrees of freedom.}
\label{tab:crossover_paired}
\small
\begin{tabular}{@{}lrrrrr@{}}
\toprule
Descriptor & Mean difference & 95\% interval & $t(2)$ & $p$ & $d_z$ \\
\midrule
Angle ($^\circ$) & 5.86 & $[0.95,10.77]$ & 5.14 & 0.036 & 2.97 \\
Count & $-3.04$ & $[-15.17,9.09]$ & $-1.08$ & 0.393 & $-0.62$ \\
\bottomrule
\end{tabular}
\end{table}

Across the three real images, angle and count are $48.72\pm2.97^\circ$ and $27.67\pm2.08$, respectively (mean $\pm$ sample SD).
Across the three synthetic per-image means, they are $54.58\pm1.40^\circ$ and $24.63\pm3.14$.
The angle SDs among synthetic variants within Images 1 to 3 are $5.51^\circ$, $6.85^\circ$, and $3.06^\circ$; the corresponding count SDs are 2.45, 2.49, and 2.83.
Thus, the SD across synthetic per-image means must not be interpreted as the variability among individual synthetic images.

\section{AFM Line-Profile Measurements Motivating Continuous Mapping}
\label{sec:s_mapping}

\begin{figure}[!htbp]
\centering
\begin{tikzpicture}
    \node[inner sep=0] (afmimages) {
        \includegraphics[
            width=0.92\linewidth
        ]{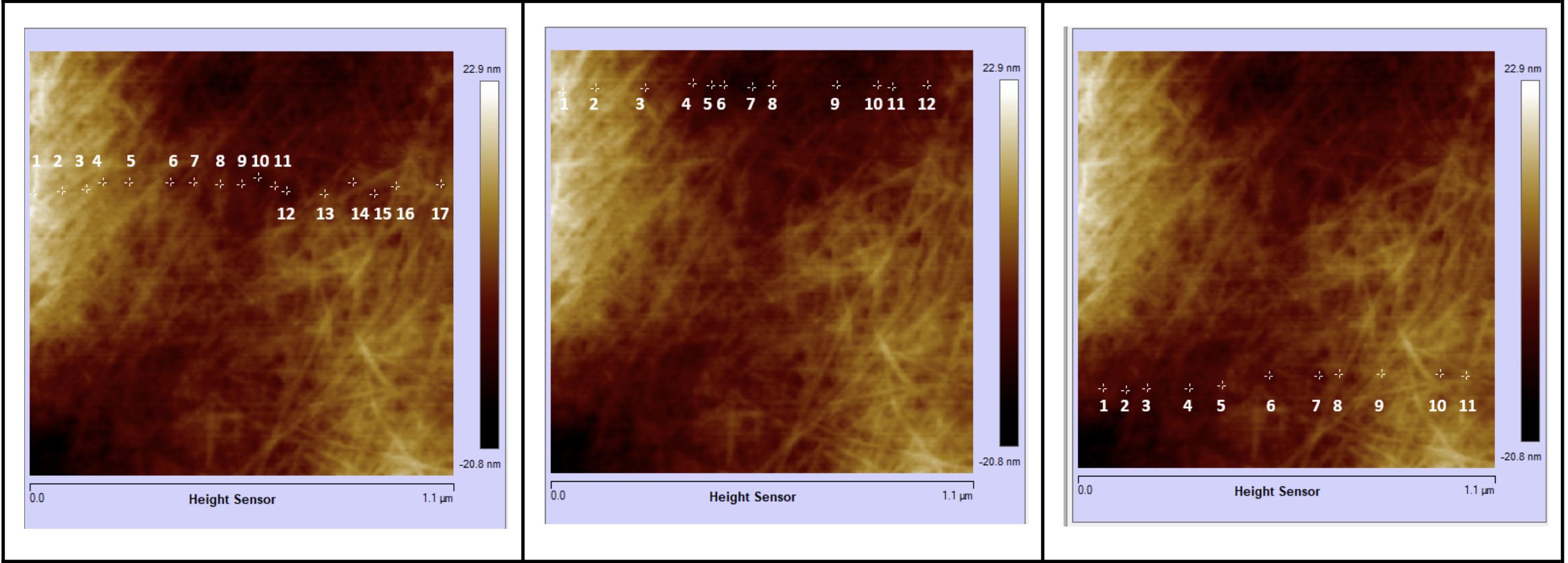}
    };
    \node[
        font=\bfseries,
        anchor=south
    ] at
    ($(afmimages.north west)!0.1667!(afmimages.north east)+(0,2mm)$)
    {AFM Image 1};
    \node[
        font=\bfseries,
        anchor=south
    ] at
    ($(afmimages.north west)!0.5000!(afmimages.north east)+(0,2mm)$)
    {AFM Image 2};
    \node[
        font=\bfseries,
        anchor=south
    ] at
    ($(afmimages.north west)!0.8333!(afmimages.north east)+(0,2mm)$)
    {AFM Image 3};
\end{tikzpicture}

\vspace{6mm}

\begin{tikzpicture}
\begin{axis}[
    width=0.92\linewidth,
    height=0.50\linewidth,
    xmin=0.5,
    xmax=17.5,
    ymin=39,
    ymax=86,
    xtick={1,2,...,17},
    xlabel={Sampling Point},
    ylabel={Young's Modulus (MPa)},
    xlabel style={
        font=\large,
        yshift=-1mm
    },
    ylabel style={
        font=\large,
        yshift=1mm
    },
    tick label style={
        font=\small
    },
    grid=major,
    major grid style={
        gray!30,
        line width=0.4pt
    },
    axis line style={
        line width=0.7pt
    },
    tick align=outside,
    mark size=2.7pt,
    line width=0.8pt,
    legend style={
        at={(0.5,1.03)},
        anchor=south,
        legend columns=3,
        draw=none,
        font=\small,
        column sep=5mm,
        /tikz/every even column/.append style={
            column sep=6mm
        }
    },
]
\addplot[
    color=blue,
    mark=o,
    mark options={
        fill=white,
        line width=0.8pt
    }
] coordinates {
    (1,78.2)
    (2,57.2)
    (3,67.3)
    (4,57.3)
    (5,56.8)
    (6,60.6)
    (7,55.7)
    (8,50.3)
    (9,50.4)
    (10,43.8)
    (11,50.2)
    (12,53.4)
    (13,70.8)
    (14,54.8)
    (15,48.2)
    (16,79.7)
    (17,46.8)
};
\addlegendentry{AFM Image 1}
\addplot[
    color=orange,
    mark=square,
    mark options={
        fill=white,
        line width=0.8pt
    }
] coordinates {
    (1,54.3)
    (2,84.2)
    (3,43.4)
    (4,59.8)
    (5,48.2)
    (6,48.6)
    (7,66.8)
    (8,47.8)
    (9,59.7)
    (10,54.1)
    (11,79.1)
    (12,57.8)
};
\addlegendentry{AFM Image 2}
\addplot[
    color=green!60!black,
    mark=triangle,
    mark options={
        fill=white,
        line width=0.8pt
    }
] coordinates {
    (1,78.9)
    (2,56.8)
    (3,67.4)
    (4,74.9)
    (5,50.7)
    (6,57.1)
    (7,60.8)
    (8,51.8)
    (9,59.7)
    (10,55.9)
    (11,44.9)
};
\addlegendentry{AFM Image 3}
\end{axis}
\end{tikzpicture}
\caption{Young's modulus measurements obtained along one-dimensional line segments extracted from three AFM images of 8~DPA cotton fibers.
    Each series gives the measured modulus, in MPa, at successive sampling points along one line segment; the series contain 17, 12, and 11 points, respectively (40 plotted measurements).}
\label{fig:afm_line_variation}
\end{figure}

Young's modulus was measured along one-dimensional line segments extracted from three AFM images (\figref{fig:afm_line_variation}).
The plotted dataset contains 40 modulus measurements: 17, 12, and 11 points in Images 1 to 3.
These data show substantial spatial heterogeneity across the sampled locations and no obvious binary separation into two phases. They do not establish a complete sampling count beyond the plotted series.
The continuous interpolation mapping of the main text retains continuous image-intensity variation in the assigned material field (\figref{fig:curveplotsMaterialProp}).
We also implemented a threshold-based mapping, in which an intensity threshold segments the image into distinct material regions and each region is assigned a representative property.
The reported results use only the continuous mapping.
The threshold-based mapping replaces the measured grayscale variability with artificially discrete regions.
Note that retaining the continuous field does not by itself validate the physical calibration between AFM intensity and local modulus.

\begin{figure}[!htbp]
    \centering
\begin{tikzpicture}
  \begin{axis}[
    width=0.95\linewidth,
    height=0.52\linewidth,
    grid=both,
    grid style={line width=0.2pt, draw=gray!25},
    major grid style={line width=0.35pt, draw=gray!35},
    xlabel={Normalized position along line},
    ylabel={Young's modulus (MPa)},
    xmin=0,
    xmax=1,
    scaled y ticks=false,
    tick label style={font=\small},
    label style={font=\small},
    legend style={
      font=\small,
      draw=none,
      fill=none,
      at={(0.5,1.02)},
      anchor=south,
      legend columns=3,
      /tikz/every even column/.append style={column sep=6mm}
    },
    cycle list={
      {mplblue, thick},
      {mplorange, thick, dashed},
      {mplgreen, thick, dash dot}
    }
  ]
    \addplot table[
      col sep=comma,
      x index=4,
      y expr=\thisrowno{2}/1e6
    ] {Figures/DataLine/data_alongline.csv};
    \addlegendentry{Image 1}
    \addplot table[
      col sep=comma,
      x index=4,
      y expr=\thisrowno{2}/1e6
    ] {Figures/DataLine/data_alongline2.csv};
    \addlegendentry{Image 2}
    \addplot table[
      col sep=comma,
      x index=4,
      y expr=\thisrowno{2}/1e6
    ] {Figures/DataLine/data_alongline3.csv};
    \addlegendentry{Image 3}
  \end{axis}
\end{tikzpicture}
\caption{Spatial variation of Young's modulus along representative line segments extracted from three AFM images after applying the continuous interpolation mapping.
    The profiles are plotted against normalized arc length along the sampled line, and they retain the pointwise heterogeneity of the AFM-derived material fields.}
\label{fig:curveplotsMaterialProp}
\end{figure}

\FloatBarrier

\section{Homogenization Formulation}
\label{sec:s_homogenization}

\subsection{Governing Equations}

Let $\bm{u}:\Omega\rightarrow\mathbb{R}^2$ denote the displacement field on the computational cell domain $\Omega$.
The boundary-value problem of heterogeneous linear elasticity reads
\begin{align}
  -\nabla\cdot\bm{\sigma}(\bm{u},\bm{x}) &= \bm{0}
    \quad \text{in } \Omega, \label{eq:equil} \\
  \bm{\sigma}(\bm{u},\bm{x}) &= \mathbb{C}(\bm{x}):\bm{\varepsilon}(\bm{u}),
    \label{eq:constitutive} \\
  \bm{\varepsilon}(\bm{u}) &= \tfrac{1}{2}
    \bigl(\nabla\bm{u} + \nabla\bm{u}^T\bigr). \label{eq:kinematics}
\end{align}
We assemble the spatially varying stiffness tensor $\mathbb{C}(\bm{x})$ at each Gauss point from the image-derived fields $E(\bm{x})$ and $\nu(\bm{x})$.
We assume plane strain throughout.
The term RVE denotes this computational cell; its statistical representativeness has not been established.

\subsection{Voigt Notation and Plane-Strain Stiffness}

In two dimensions we adopt the standard Voigt (engineering-shear) convention for the strain and stress vectors,
\begin{equation}
  \bm{\varepsilon}^V(\bm{u})
  =
  \begin{bmatrix}
    \varepsilon_{xx} \\ \varepsilon_{yy} \\ \gamma_{xy}
  \end{bmatrix}
  =
  \begin{bmatrix}
    u_{x,x} \\ u_{y,y} \\ u_{x,y}+u_{y,x}
  \end{bmatrix},
  \qquad
  \bm{\sigma}^V
  =
  \begin{bmatrix}
    \sigma_{xx} \\ \sigma_{yy} \\ \tau_{xy}
  \end{bmatrix},
\end{equation}
and the constitutive relation at each point becomes $\bm{\sigma}^V(\bm{x}) = C(\bm{x})\,\bm{\varepsilon}^V(\bm{x})$, with $C(\bm{x})\in\mathbb{R}^{3\times 3}$.
For an isotropic material under plane strain the local stiffness matrix is
\begin{equation}
  C_{\mathrm{pe}}
  =
  \begin{bmatrix}
    \dfrac{E(1-\nu)}{(1+\nu)(1-2\nu)} &
    \dfrac{E\nu}{(1+\nu)(1-2\nu)} & 0 \\[8pt]
    \dfrac{E\nu}{(1+\nu)(1-2\nu)} &
    \dfrac{E(1-\nu)}{(1+\nu)(1-2\nu)} & 0 \\[8pt]
    0 & 0 & \dfrac{E}{2(1+\nu)}
  \end{bmatrix}.
  \label{eq:Cpe}
\end{equation}

\subsection{Macroscopic Load Cases and Boundary Conditions}

We prescribe three linearly independent macroscopic strain states to identify all columns of the effective stiffness,
\begin{equation}
  \bar{\bm{\varepsilon}}^{(A)}
  = \begin{bmatrix}1\\0\\0\end{bmatrix}, \qquad
  \bar{\bm{\varepsilon}}^{(B)}
  = \begin{bmatrix}0\\1\\0\end{bmatrix}, \qquad
  \bar{\bm{\varepsilon}}^{(C)}
  = \begin{bmatrix}0\\0\\1\end{bmatrix},
  \label{eq:loadcases}
\end{equation}
corresponding to unit longitudinal strain $\varepsilon_{xx}$, unit transverse strain $\varepsilon_{yy}$, and unit engineering shear $\gamma_{xy}$.
We enforce each load case through affine displacement boundary conditions on $\partial\Omega$,
\begin{equation}
  \bm{u}^{(k)}(\bm{x}) = \bar{\boldsymbol{\epsilon}}^{(k)}\bm{x}
  \quad \text{on } \partial\Omega, \quad k\in\{A,B,C\},
  \label{eq:affineBCs}
\end{equation}
where $\bar{\boldsymbol{\epsilon}}^{(k)}$ is the $2\times2$ strain tensor corresponding to the engineering-strain vector in \eqnref{eq:loadcases},
\begin{equation}
\bar{\boldsymbol{\epsilon}}^{(k)}=
\begin{bmatrix}
\bar{\varepsilon}_{xx}^{(k)} & \bar{\gamma}_{xy}^{(k)}/2\\
\bar{\gamma}_{xy}^{(k)}/2 & \bar{\varepsilon}_{yy}^{(k)}
\end{bmatrix}.
\end{equation}
We use the normalized coordinates $\hat{x}=(x-x_0)/L_x$ and $\hat{y}=(y-y_0)/L_y$ in the implementation.
This keeps all boundary displacements of order one.
We perform all homogenization on a unit RVE ($L_x = L_y = 1$) in non-dimensional units.
The normalized coordinates then coincide with the computational coordinates, and each load case imposes a strain of unit magnitude.
This unit-domain calculation does not establish the physical size or representativeness of the image window.
The explicit forms are
\begin{align}
  \text{Case A:} & \quad u_x = \hat{x},\quad u_y = 0, \\
  \text{Case B:} & \quad u_x = 0,\quad u_y = \hat{y}, \\
  \text{Case C:} & \quad u_x = \tfrac{1}{2}\hat{y},\quad u_y = \tfrac{1}{2}\hat{x}.
\end{align}

\subsection{Volume-Averaged Stress and Effective Stiffness}

After solving each sub-problem, we compute the volume-averaged (macroscopic) stress as
\begin{equation}
  \langle\bm{\sigma}^V\rangle^{(k)}
  := \frac{1}{|\Omega|}
     \int_{\Omega}\bm{\sigma}^V\!\bigl(\bm{u}^{(k)}(\bm{x}),\bm{x}\bigr)\,
     \mathrm{d}\Omega,
  \qquad |\Omega| = \int_{\Omega}1\,\mathrm{d}\Omega,
  \label{eq:avgstress}
\end{equation}
evaluated in the finite element setting by summing over all elements $e$ and Gauss points $g$,
\begin{equation}
  \langle\bm{\sigma}^V\rangle^{(k)}
  \approx
  \frac{\displaystyle\sum_{e}\sum_{g}
        \bm{\sigma}^V_{eg}\,w_{eg}}
       {\displaystyle\sum_{e}\sum_{g} w_{eg}},
  \qquad w_{eg} = \det J_{eg}\,w^{\mathrm{quad}}_g,
  \label{eq:femavg}
\end{equation}
where $J_{eg}$ is the Jacobian of the isoparametric mapping and $w^{\mathrm{quad}}_g$ are the quadrature weights.
The homogenized constitutive law is $\langle\bm{\sigma}^V\rangle = C_{\mathrm{eff}}\,\bar{\bm{\varepsilon}}$.
The three unit load cases in \eqnref{eq:loadcases} form the identity matrix.
The columns of $C_{\mathrm{eff}}$ then follow immediately,
\begin{equation}
  C_{\mathrm{eff}}
  =
  \begin{bmatrix}
    \langle\sigma_{xx}\rangle^{(A)} &
    \langle\sigma_{xx}\rangle^{(B)} &
    \langle\sigma_{xx}\rangle^{(C)} \\
    \langle\sigma_{yy}\rangle^{(A)} &
    \langle\sigma_{yy}\rangle^{(B)} &
    \langle\sigma_{yy}\rangle^{(C)} \\
    \langle\tau_{xy}\rangle^{(A)} &
    \langle\tau_{xy}\rangle^{(B)} &
    \langle\tau_{xy}\rangle^{(C)}
  \end{bmatrix}.
  \label{eq:Ceff}
\end{equation}
When we apply a load case with amplitude $\varepsilon_0$ (or $\gamma_0$) instead of unity, we divide the corresponding column of \eqnref{eq:Ceff} by that amplitude.

\subsection{Isotropy Check and Extraction of Effective Constants}
\label{sec:s_isotropy}

An in-plane isotropic effective stiffness must satisfy
\begin{equation}
  C_{11}\approx C_{22}, \qquad
  C_{13}\approx C_{23}\approx 0, \qquad
  2C_{33}\approx C_{11}-C_{12}.
  \label{eq:isotropy_check}
\end{equation}
Reciprocity also requires $C_{ij}\approx C_{ji}$, but this symmetry alone does not establish isotropy.

\subsubsection*{Isotropy screen applied to the ensemble}

After assembling $C_{\mathrm{eff}}$, the solver screens every run for approximate isotropy.
With the relative difference
\begin{equation}
r(a,b)=\frac{|a-b|}{\max\left(1,|a|,|b|\right)},
\label{eq:reldiff}
\end{equation}
where the unit floor in the denominator is a numerical safeguard that is negligible at the stiffness magnitudes of this study, a run is labeled approximately isotropic only if all four inequalities hold,
\begin{equation}
r(C_{12},C_{21})<0.01,\qquad
r(C_{11},C_{22})<0.01,\qquad
|C_{13}|<0.01\,|C_{11}|,\qquad
|C_{23}|<0.01\,|C_{11}|,
\label{eq:iso_screen}
\end{equation}
and anisotropic otherwise.
The first inequality tests reciprocity, whose failure would indicate a numerical or consistency problem, not physical anisotropy.
The second tests equality of the normal stiffnesses along the image axes, and the last two bound the normal-shear coupling.
We report the normal-stiffness criterion as the continuous directional mismatch
\begin{equation}
A_N=100\,r(C_{11},C_{22})\ (\%),
\label{eq:normal_mismatch}
\end{equation}
which the screen requires to be below 1\%.
$A_N$ measures directional normal-stiffness mismatch only and is not a complete anisotropy index.

\tabref{tab:anisotropy} summarizes the screen over the 798 completed homogenization runs.
All runs pass the reciprocity check.
Every run at $E_m/E_f=0.91$, 1.45, and 1.81 passes all four criteria.
At $E_m/E_f=0.18$, 24 of 200 runs (12.0\%) are labeled anisotropic, and all 24 fail the normal-stiffness criterion, with $A_N$ up to 1.30\%.
These cases are concentrated in three of the five parent images, which contribute 11, 11, and 2 of their 40 variants, while the other two parent images contribute none.
The saved logs do not establish whether any of these 24 runs also fail a coupling criterion.
Pooled over all four ratios, 24 of 798 runs (3.0\%) are labeled anisotropic, with median $A_N = 0.33\%$ and maximum $A_N = 1.30\%$.
Because the same microstructures recur at every stiffness ratio, the pooled percentage is a run-level description, not a sample of independent microstructures.
The mean mismatch is smallest when matrix and fibril stiffnesses are comparable ($E_m/E_f=0.91$ and 1.45) and increases with the stiffness contrast between the phases, reaching its largest value for the most compliant matrix ($E_m/E_f=0.18$).

\begin{table}[!htbp]
\centering
\caption{Isotropy screen of the homogenized stiffness tensors for each matrix-to-fibril stiffness ratio. A run is anisotropic if it fails any criterion in \eqnref{eq:iso_screen}. $A_N$ is the normal-stiffness mismatch of \eqnref{eq:normal_mismatch}; SD is the sample standard deviation and P95 the 95th percentile. Two failed runs without a homogenization summary are excluded.}
\label{tab:anisotropy}
\small
\setlength{\tabcolsep}{4pt}
\begin{tabular}{@{}lrrrrrrrr@{}}
\toprule
 & & \multicolumn{2}{c}{Anisotropic} & \multicolumn{5}{c}{$A_N$ (\%)} \\
\cmidrule(lr){3-4}\cmidrule(l){5-9}
$E_m/E_f$ & Valid runs & $n$ & \% & Mean & SD & Median & P95 & Max \\
\midrule
0.18 & 200 & 24 & 12.0 & 0.820 & 0.156 & 0.824 & 1.073 & 1.300 \\
0.91 & 199 & 0 & 0.0 & 0.236 & 0.048 & 0.237 & 0.319 & 0.362 \\
1.45 & 199 & 0 & 0.0 & 0.252 & 0.053 & 0.256 & 0.341 & 0.391 \\
1.81 & 200 & 0 & 0.0 & 0.494 & 0.105 & 0.501 & 0.663 & 0.767 \\
\midrule
All  & 798 & 24 & 3.0 & 0.451 & 0.257 & 0.334 & 0.958 & 1.300 \\
\bottomrule
\end{tabular}
\end{table}

The screen has two limitations.
First, it does not test the isotropic shear relation $2C_{33}=C_{11}-C_{12}$ of \eqnref{eq:isotropy_check}, whose residual would be
\begin{equation}
\eta_S=\frac{|2C_{33}-(C_{11}-C_{12})|}{|C_{11}-C_{12}|}.
\label{eq:shear_residual}
\end{equation}
Passing the screen therefore indicates small normal-stiffness mismatch, reciprocity, and small normal-shear coupling, but does not by itself establish full in-plane isotropy.
Second, the solver logs record the relative mismatches and the coupling terms but not the complete tensor, so $\eta_S$ and full-tensor anisotropy indices cannot be recovered for the present ensemble. Saving all entries of $C_{\mathrm{eff}}$ in future runs would remove this limitation.
Deviations from isotropy can also reflect finite-window effects or numerical error, and the screen alone cannot identify their cause.

\subsubsection*{Effective constants}

Under an isotropic reduction, the effective Young's modulus and Poisson's ratio follow from the plane-strain identities
\begin{equation}
  \nu_{\mathrm{eff}}
  = \frac{C_{12}}{C_{11}+C_{12}},
  \qquad
  E_{\mathrm{eff}}
  = 2\,C_{33}\,(1+\nu_{\mathrm{eff}}).
  \label{eq:extract_pe}
\end{equation}
These relations follow from inverting the plane-strain stiffness in \eqnref{eq:Cpe} and are exact for a homogeneous isotropic material.
Applied to the heterogeneous effective tensor, they define the scalar approximation used here, whose accuracy depends on the isotropy criteria above.
A consistency check is the independently reconstructed normal-stiffness expression
\begin{equation}
E_{\mathrm{eff}}^{(11)}=
C_{11}\frac{(1+\nu_{\mathrm{eff}})(1-2\nu_{\mathrm{eff}})}{1-\nu_{\mathrm{eff}}}.
\end{equation}
It equals $2C_{33}(1+\nu_{\mathrm{eff}})$ only when the relevant isotropic identities hold.
No numerical comparison of these expressions is reported here.

\section{Macroscale Assembly with Sampled Properties}
\label{sec:s_upscaling}

The homogenization ensemble provides paired effective properties $\mathcal{S} = \{(E_{\mathrm{eff}}^{(j)}, \nu_{\mathrm{eff}}^{(j)})\}_{j=1}^{N_s}$, whose empirical joint density is
\begin{equation}
\widehat{p}_{E,\nu}(E,\nu) = \frac{1}{N_s} \sum_{j=1}^{N_s} \delta\!\left(E-E_{\mathrm{eff}}^{(j)}\right) \delta\!\left(\nu-\nu_{\mathrm{eff}}^{(j)}\right),
\end{equation}
where $\delta(\cdot)$ denotes the Dirac delta distribution.
We draw an index $J_g \sim \mathcal{U}\{1,\ldots,N_s\}$ at each Gauss point $\mathbf{x}_g$ and assign $(E(\mathbf{x}_g), \nu(\mathbf{x}_g)) = (E_{\mathrm{eff}}^{(J_g)}, \nu_{\mathrm{eff}}^{(J_g)})$.
This samples the density directly.
The local constitutive tensor $\mathbb{C}(\mathbf{x}_g) = \mathbb{C}(E(\mathbf{x}_g), \nu(\mathbf{x}_g))$ then enters the weak form of the macroscale linear-elasticity problem, whose internal virtual work is approximated as
\begin{equation}
\int_{\Omega} \boldsymbol{\varepsilon}(\mathbf{v}) : \mathbb{C}(\mathbf{x}) : \boldsymbol{\varepsilon}(\mathbf{u}) \, \mathrm{d}\Omega
\approx
\sum_e \sum_g w_g\, J_g^{\mathrm{det}}\, \boldsymbol{\varepsilon}(\mathbf{v}_h)(\mathbf{x}_g) : \mathbb{C}(\mathbf{x}_g) : \boldsymbol{\varepsilon}(\mathbf{u}_h)(\mathbf{x}_g),
\end{equation}
where $w_g$ is the quadrature weight and $J_g^{\mathrm{det}}$ denotes the determinant of the element mapping Jacobian at the Gauss point, which is a different quantity from the random sample index $J_g$.

\section{Additional Results}
\label{sec:s_results}

\subsection{Displacement Fields from the Homogenization Load Cases}

\begin{figure}[!htbp]
    \centering
    \captionsetup[subfigure]{font=small,labelfont=bf,skip=2pt}
    \begin{subfigure}[t]{0.32\linewidth}
        \centering
        \pvzoompanel{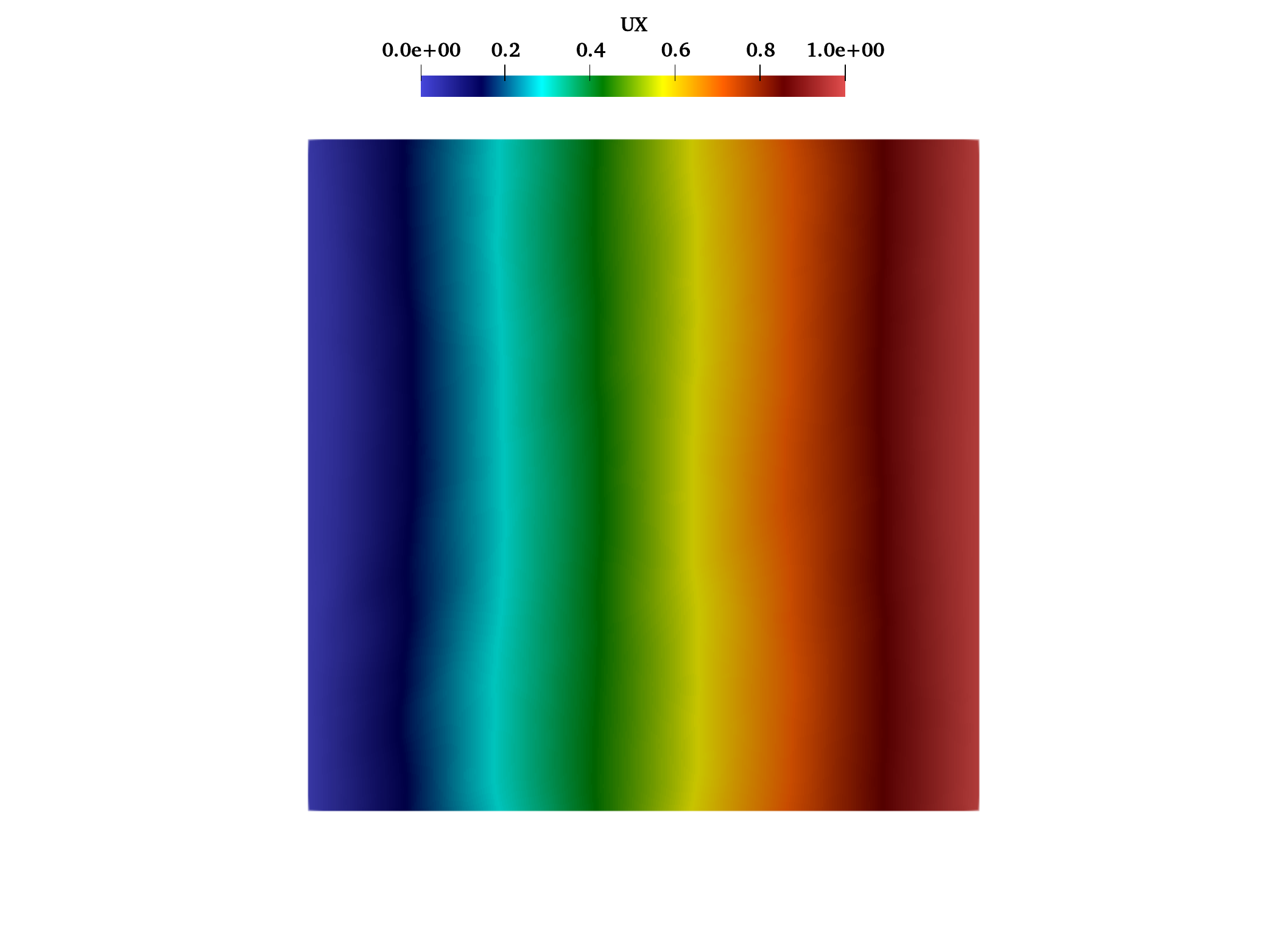}
        \caption{Case~A, $u_x$.}
        \label{fig:case_A_ux}
    \end{subfigure}
    \hfill
    \begin{subfigure}[t]{0.32\linewidth}
        \centering
        \pvzoompanel{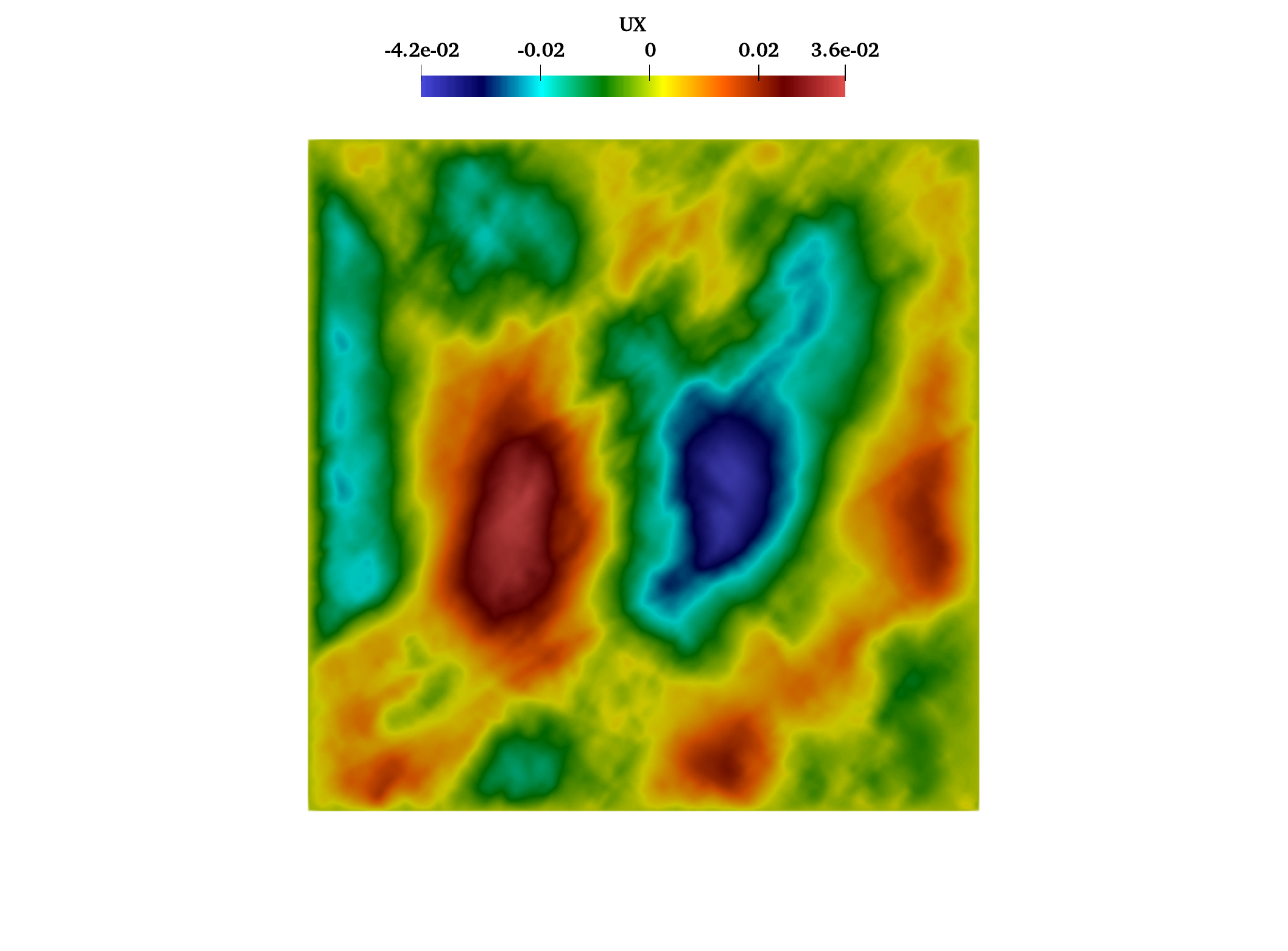}
        \caption{Case~B, $u_x$.}
        \label{fig:case_B_ux}
    \end{subfigure}
    \hfill
    \begin{subfigure}[t]{0.32\linewidth}
        \centering
        \pvzoompanel{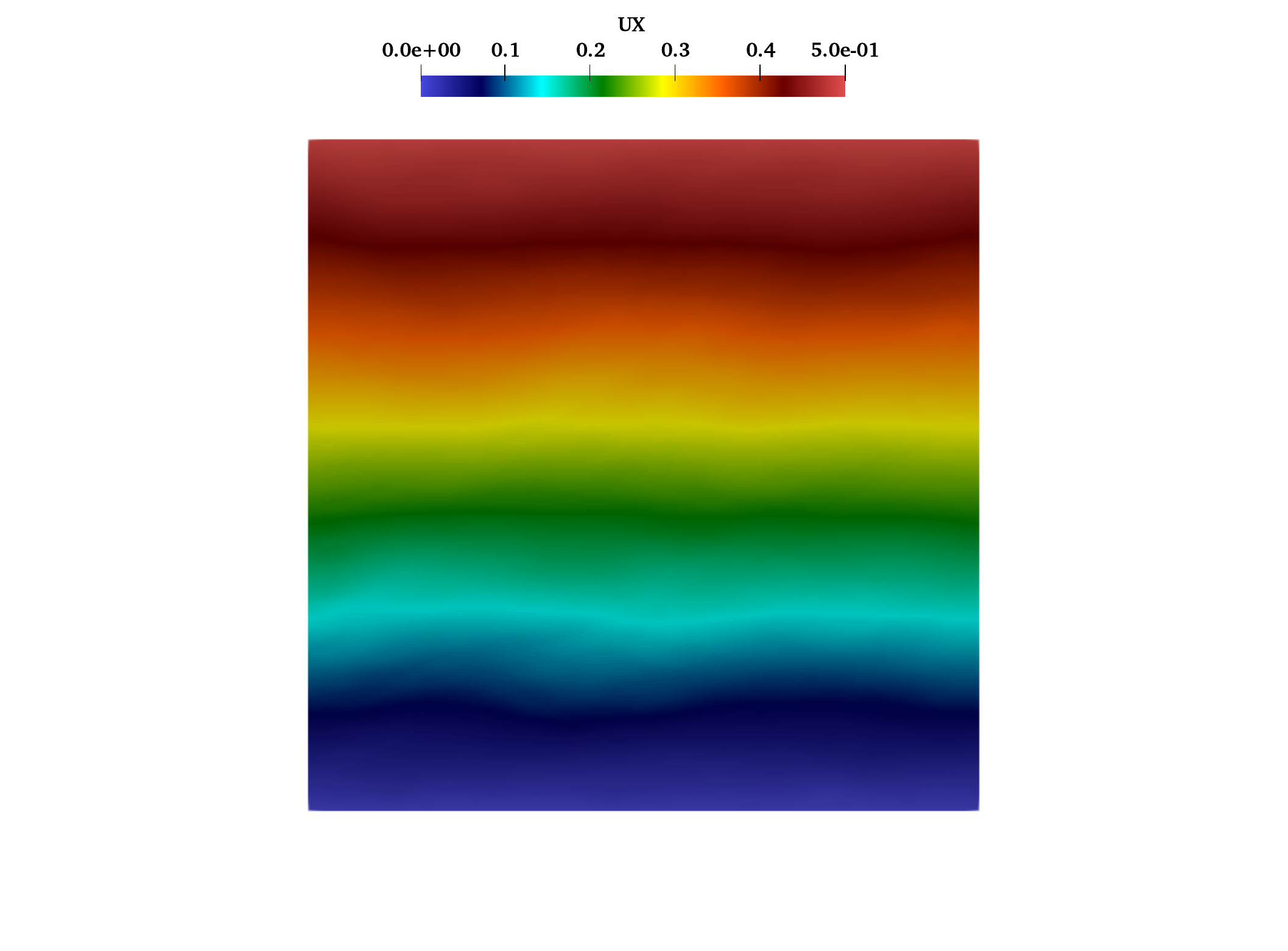}
        \caption{Case~C, $u_x$.}
        \label{fig:case_C_ux}
    \end{subfigure}

    \vspace{0.6em}

    \begin{subfigure}[t]{0.32\linewidth}
        \centering
        \pvzoompanel{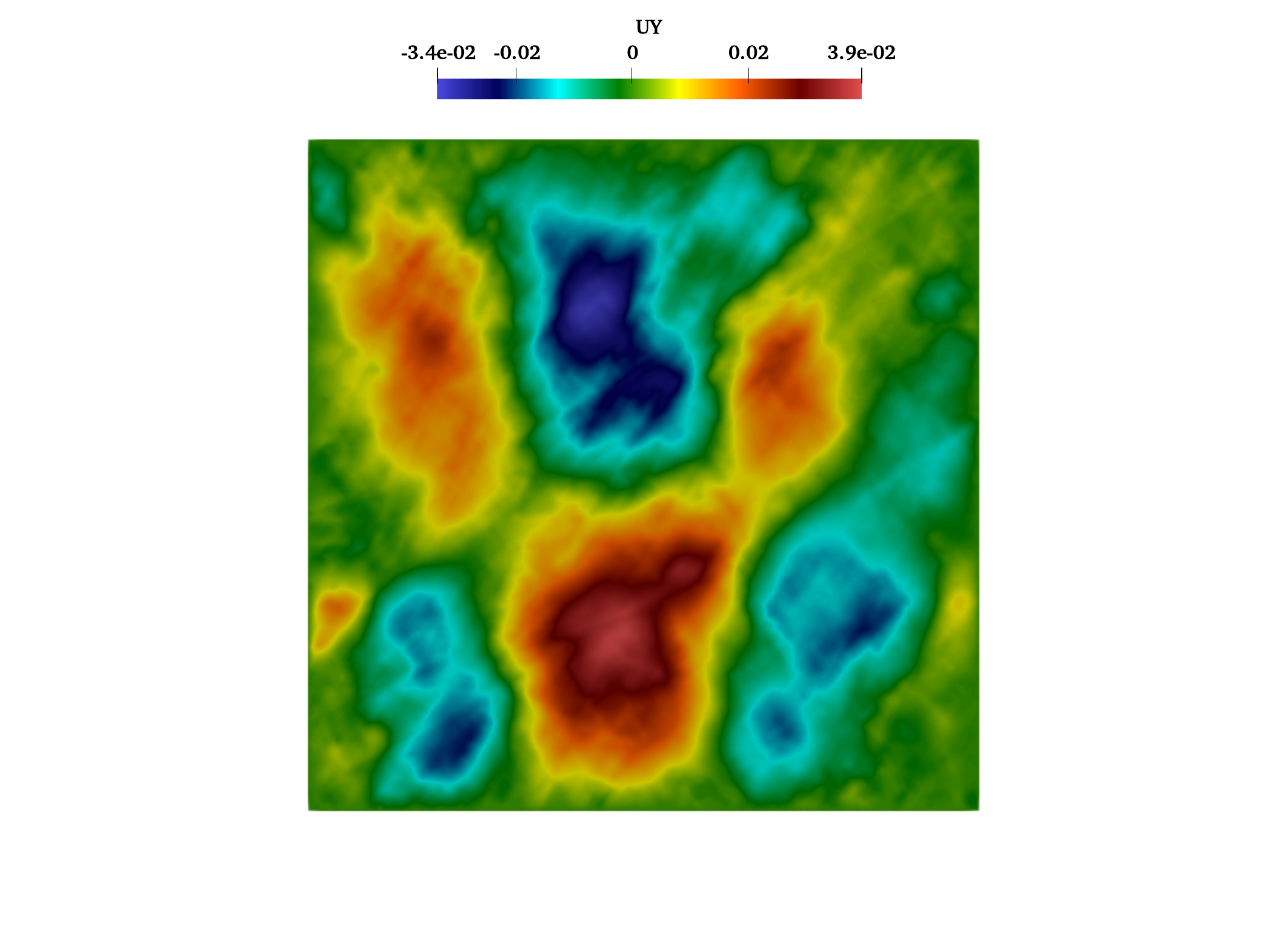}
        \caption{Case~A, $u_y$.}
        \label{fig:case_A_uy}
    \end{subfigure}
    \hfill
    \begin{subfigure}[t]{0.32\linewidth}
        \centering
        \pvzoompanel{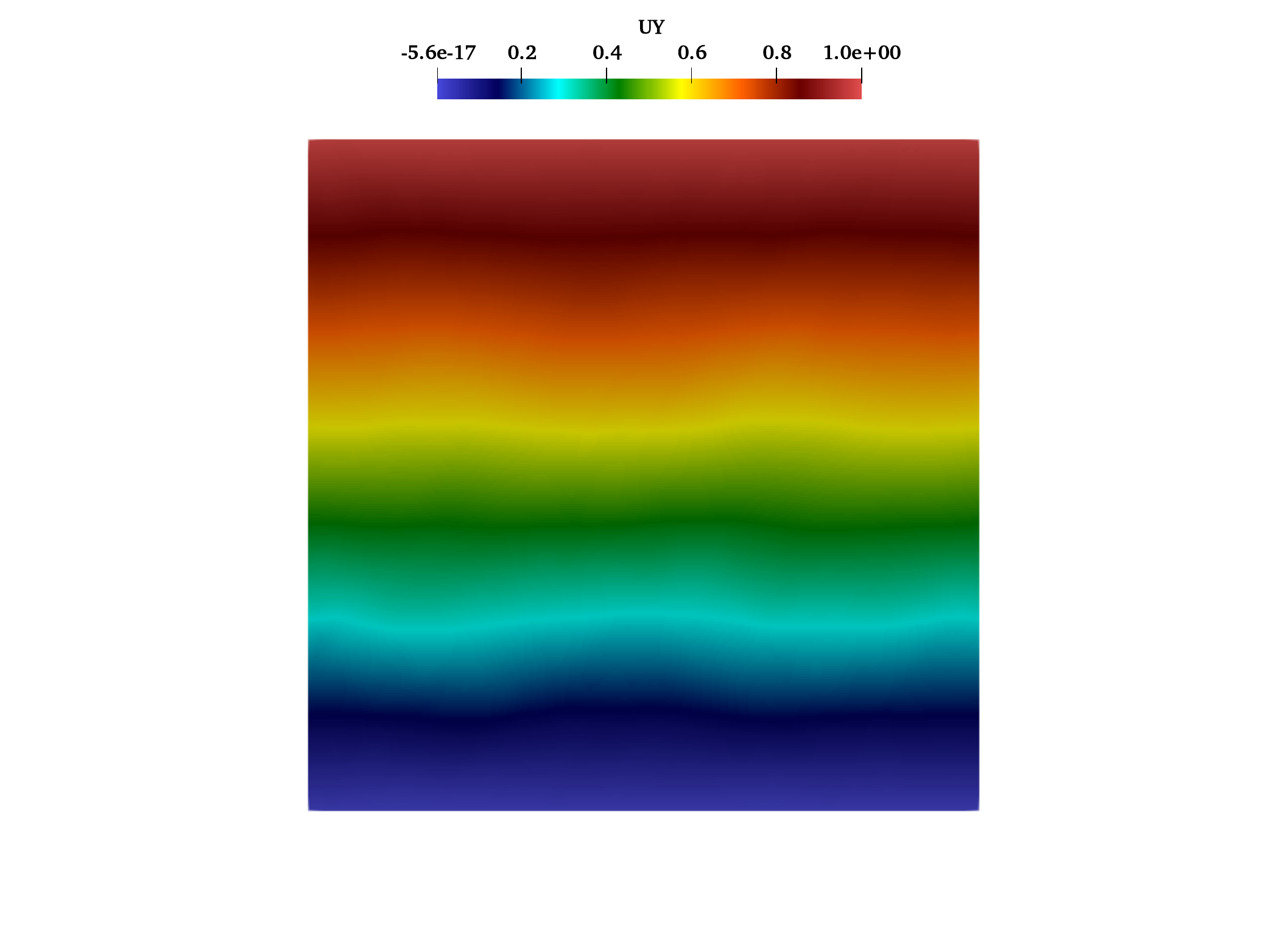}
        \caption{Case~B, $u_y$.}
        \label{fig:case_B_uy}
    \end{subfigure}
    \hfill
    \begin{subfigure}[t]{0.32\linewidth}
        \centering
        \pvzoompanel{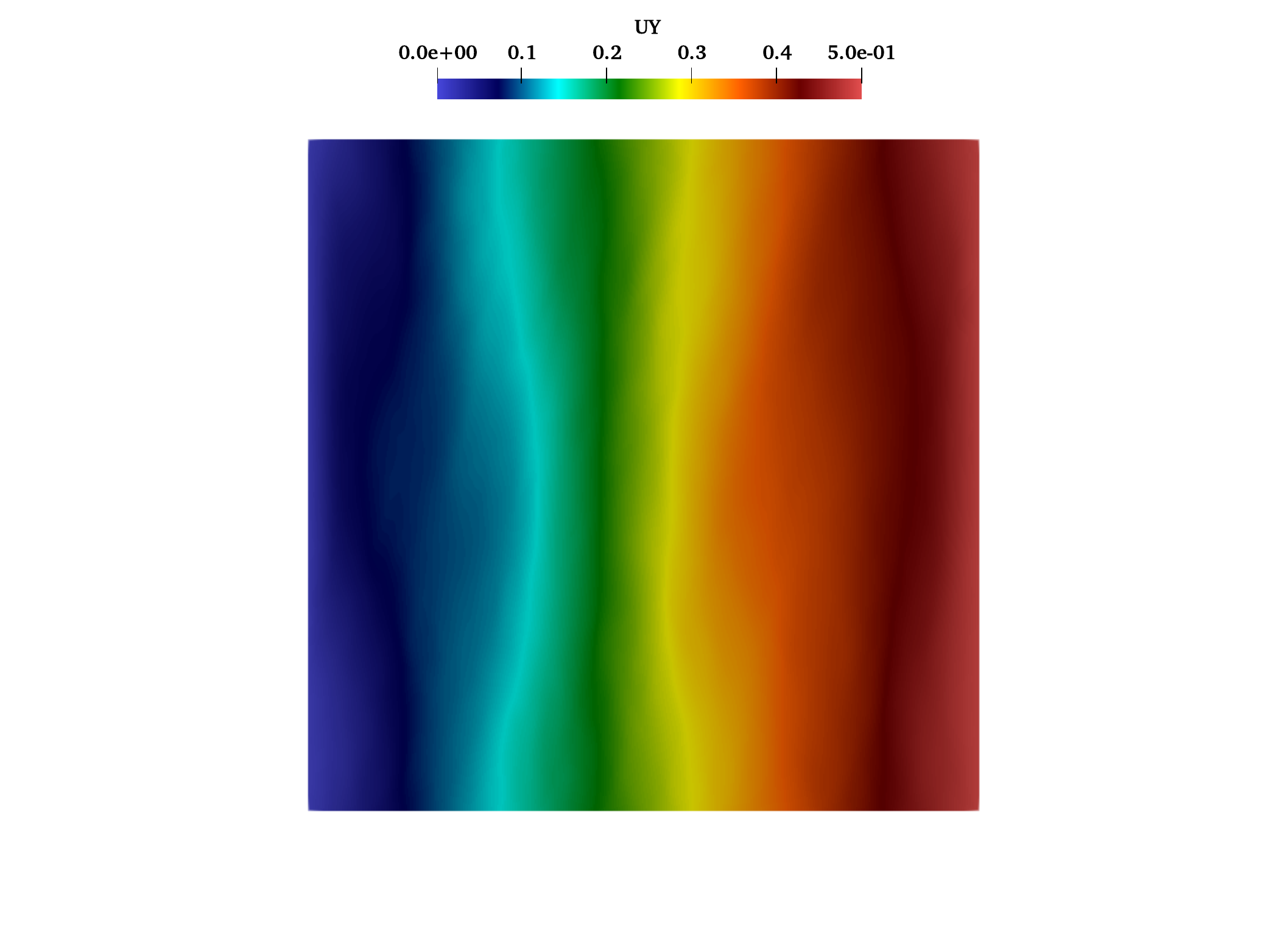}
        \caption{Case~C, $u_y$.}
        \label{fig:case_C_uy}
    \end{subfigure}
    \caption{Displacement fields obtained from the three independent macroscopic strain loading cases used for computational homogenization.
        Case~A corresponds to unit longitudinal strain $(\bar{\bm{\varepsilon}} = [1,0,0]^T)$, Case~B corresponds to unit transverse strain $(\bar{\bm{\varepsilon}} = [0,1,0]^T)$, and Case~C corresponds to unit engineering shear $(\bar{\bm{\varepsilon}} = [0,0,1]^T,\ \gamma_{xy}=1)$.}
    \label{fig:homogenization_displacement_cases}
\end{figure}

\figref{fig:homogenization_displacement_cases} shows the displacement fields $u_x$ and $u_y$ for each of the three macroscopic load cases applied to a representative RVE.
For Case~A (unit longitudinal strain), the $u_x$ field follows the imposed left-to-right gradient, with local perturbations associated with property heterogeneity.
The $u_y$ field for the same case contains transverse in-plane displacement arising from the Poisson response and local heterogeneity.
Case~B (unit transverse strain) produces the complementary pattern, with the primary gradient in $u_y$ and the transverse response in $u_x$.
Case~C (unit engineering shear) produces spatial variation in both displacement components.
Even under nominally uniform macroscopic strains, local stiffness contrasts drive substantial fluctuations in the microscale displacement field.
These fluctuations enter the volume-averaged stresses used to construct $C_{\mathrm{eff}}$.

\subsection{Effective-Modulus Density for a Single Matrix Stiffness}

\begin{figure}[!htbp]
    \centering
    \includegraphics[width=0.95\linewidth]{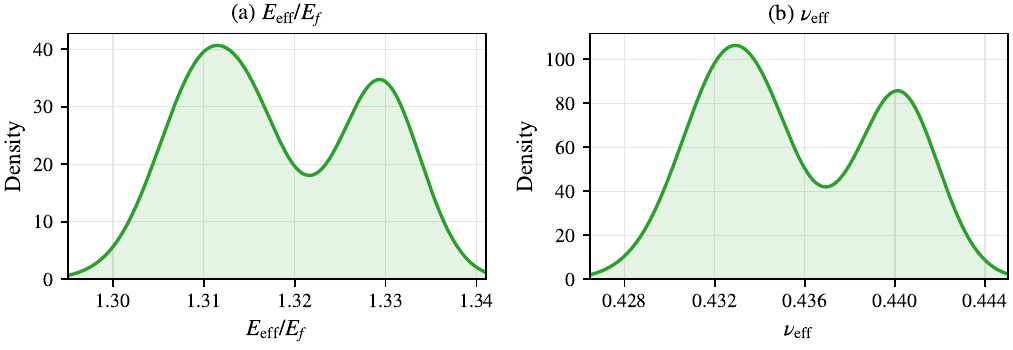}
    \caption{Marginal kernel density estimates of $E_{\mathrm{eff}}/E_f$ and $\nu_{\mathrm{eff}}$ for $E_m/E_f = 1.45$ ($E_m = 80$~MPa).
        The effective-modulus density contains two visible peaks separated by a lower-density region.
        Parent-image grouping and KDE bandwidth require assessment before interpreting these peaks as distinct physical populations.}
    \label{fig:pdf_80}
\end{figure}

\figref{fig:pdf_80} shows the marginal kernel density estimates (KDEs) of the normalized effective modulus $E_{\mathrm{eff}}/E_f$ and of $\nu_{\mathrm{eff}}$ for $E_m/E_f = 1.45$ ($E_m = 80$~MPa, $E_f = 55.1$~MPa).
The modulus marginal is considered here in isolation.
The density contains two visible peaks separated by a lower-density region.
We have not tested how these apparent modes depend on the KDE bandwidth.
This is left as future work.

\subsection{Mesh Convergence of the Homogenized Constants}
\label{sec:s_convergence}

We assessed the sensitivity of the homogenized constants to mesh resolution by repeating the homogenization of one representative mapped image at uniform refinement levels $L=5$ to $9$.
The unit-square RVE is discretized with bilinear (Q1) quadrilateral elements of size $h=2^{-L}$ under plane strain.
Only the base refinement level was changed between runs; the image, the continuous mapping ($E_m=10$~MPa, $\nu_m=0.49$; $E_f=55.1$~MPa, $\nu_f=0.30$), and the direct solver settings (relative tolerance $10^{-8}$) were held fixed.
The effective constants are extracted with $\nu_{\mathrm{eff}}=C_{12}/(C_{11}+C_{12})$ and $E_{\mathrm{eff}}=2C_{33}(1+\nu_{\mathrm{eff}})$ (\secref{sec:s_homogenization}).

\tabref{tab:convergence} and \figref{fig:convergence} report the results. $E_{\mathrm{eff}}$ decreases monotonically with refinement, and its difference from the $L=9$ value falls from 1.57\% at $L=5$ to 0.076\% at $L=8$.
The Poisson's ratio changes by less than 0.15\% over the full range, and the $L=8$ value differs from the $L=9$ value by 0.016\%.
Differences are computed as $100\,|Q_L-Q_9|/|Q_9|$.
The $L=9$ mesh is a finite-resolution reference, not an exact solution.
These results support mesh stability of the volume-averaged constants over the tested levels; they do not establish convergence of local stress fields.

\begin{table}[!htbp]
\centering
\caption{Mesh convergence of the homogenized constants for one representative image with $E_m=10$~MPa. Differences are relative to the $L=9$ mesh.}
\label{tab:convergence}
\begin{tabular}{rrrcccc}
\toprule
$L$ & Elements & Nodes & $E_{\mathrm{eff}}$ (MPa) & $\nu_{\mathrm{eff}}$ & $\Delta E_{\mathrm{eff}}$ (\%) & $\Delta \nu_{\mathrm{eff}}$ (\%) \\
\midrule
5 & 1024   & 1089   & 21.3700 & 0.445098 & 1.5675 & 0.1427 \\
6 & 4096   & 4225   & 21.2301 & 0.445179 & 0.9026 & 0.1245 \\
7 & 16384  & 16641  & 21.1080 & 0.445489 & 0.3222 & 0.0550 \\
8 & 65536  & 66049  & 21.0562 & 0.445662 & 0.0760 & 0.0162 \\
9 & 262144 & 263169 & 21.0402 & 0.445734 & ---    & ---    \\
\bottomrule
\end{tabular}
\end{table}

\begin{figure}[!htbp]
    \centering
    \begin{tikzpicture}
\pgfplotsset{
    convstyle/.style={
        width=0.31\linewidth,
        height=0.23\linewidth,
        scale only axis=true,
        tick label style={font=\footnotesize},
        label style={font=\small},
        tick align=inside,
        every axis plot/.append style={line width=0.9pt, mark size=1.8pt},
    },
}
\definecolor{convE}{HTML}{1F4E9C}
\definecolor{convNu}{HTML}{C0392B}

\begin{axis}[
    convstyle,
    name=conva,
    xmin=4.7, xmax=9.3,
    xtick={5,6,7,8,9},
    ymin=21.0, ymax=21.4,
    ytick={21.0,21.1,21.2,21.3,21.4},
    yticklabel style={/pgf/number format/fixed,/pgf/number format/fixed zerofill,/pgf/number format/precision=1,text=convE},
    xlabel={Refinement level $L$ ($h=2^{-L}$)},
    ylabel={$E_{\mathrm{eff}}$ (MPa)},
    ylabel style={text=convE},
    axis y line*=left,
    title={(a)}, title style={at={(0,1)}, anchor=south west, font=\small\bfseries},
]
\addplot[convE, mark=*] coordinates {
    (5,21.3700) (6,21.2301) (7,21.1080) (8,21.0562) (9,21.0402)
};
\end{axis}
\begin{axis}[
    convstyle,
    at={(conva.south west)}, anchor=south west,
    xmin=4.7, xmax=9.3,
    axis x line=none,
    ymin=0.4450, ymax=0.4458,
    ytick={0.4450,0.4452,0.4454,0.4456,0.4458},
    yticklabel style={/pgf/number format/fixed,/pgf/number format/fixed zerofill,/pgf/number format/precision=4,text=convNu},
    scaled y ticks=false,
    ylabel={$\nu_{\mathrm{eff}}$},
    ylabel style={text=convNu},
    axis y line*=right,
    legend style={font=\footnotesize, draw=none, fill=none, at={(0.97,0.50)}, anchor=east},
    legend cell align=left,
]
\addlegendimage{convE, mark=*, line width=0.9pt, mark size=1.8pt}\addlegendentry{$E_{\mathrm{eff}}$}
\addplot[convNu, mark=square*, dashed] coordinates {
    (5,0.445098) (6,0.445179) (7,0.445489) (8,0.445662) (9,0.445734)
};
\addlegendentry{$\nu_{\mathrm{eff}}$}
\end{axis}

\begin{axis}[
    convstyle,
    at={($(conva.south east)+(0.22\linewidth,0)$)}, anchor=south west,
    ymode=log,
    xmin=4.7, xmax=8.3,
    xtick={5,6,7,8},
    ymin=0.01, ymax=3,
    xlabel={Refinement level $L$},
    ylabel={Difference from $L=9$ (\%)},
    grid=major,
    major grid style={gray!25, line width=0.3pt},
    legend style={font=\footnotesize, draw=none, fill=none, at={(0.03,0.03)}, anchor=south west},
    legend cell align=left,
    title={(b)}, title style={at={(0,1)}, anchor=south west, font=\small\bfseries},
]
\addplot[convE, mark=*] coordinates {
    (5,1.5675) (6,0.9026) (7,0.3222) (8,0.0760)
};
\addlegendentry{$E_{\mathrm{eff}}$}
\addplot[convNu, mark=square*, dashed] coordinates {
    (5,0.1427) (6,0.1245) (7,0.0550) (8,0.0162)
};
\addlegendentry{$\nu_{\mathrm{eff}}$}
\end{axis}
\end{tikzpicture}
    \caption{Mesh convergence of the homogenized constants for one representative image ($E_m=10$~MPa).
        (a)~$E_{\mathrm{eff}}$ (left axis, solid) and $\nu_{\mathrm{eff}}$ (right axis, dashed) versus refinement level $L$, with element size $h=2^{-L}$.
        (b)~Relative difference from the $L=9$ reference on a logarithmic scale.
        The $L=9$ point is omitted from (b) because its difference is zero by definition.}
    \label{fig:convergence}
\end{figure}

\bibliographystyle{elsarticle-harv}
\bibliography{Final}